\documentclass[journal]{IEEEtran}

\usepackage{cite}
\usepackage{graphicx}
\usepackage{amsmath}
\usepackage{algorithmic}
\usepackage{booktabs}
\usepackage{array}
\usepackage{stfloats}
\usepackage{amssymb}
\usepackage{textcomp}
\usepackage[font=footnotesize]{caption}
\usepackage{siunitx}
\usepackage{soul}
\usepackage{xcolor}

\usepackage{xifthen}
\newcounter{reviewer}
\newcounter{point}[reviewer]
\renewcommand{\thepoint}{P\,\thereviewer.\arabic{point}}

\newcommand{\shortreply}[2][]{\medskip \noindent \begin{sf}\textbf{Reply}:\ #2
 \ifthenelse{\equal{#1}{}}{}{ \hfill \footnotesize (#1)}%
 \medskip \end{sf}}

\newcommand{\fix}[2]{{#1}\textcolor{black}{#2}}

\DeclareGraphicsExtensions{.pdf,.jpeg,.png}

\graphicspath{{.}{figures/}{figuresN/}}

\PassOptionsToPackage{draft}{hyperref} 

\begin{document}

\title{Performance-Aware Predictive-Model-Based On-Chip Body-Bias Regulation Strategy for an ULP Multi-Core Cluster in 28nm UTBB FD-SOI}


\author{


\IEEEauthorblockN{  
Alfio Di Mauro\IEEEauthorrefmark{2},
                    Davide Rossi\IEEEauthorrefmark{1}, 
                    Antonio Pullini\IEEEauthorrefmark{2}, 
                    Philippe Flatresse\IEEEauthorrefmark{3}, 
                    Luca Benini\IEEEauthorrefmark{1}\IEEEauthorrefmark{2}}
                    
\IEEEauthorblockA{  
\IEEEauthorrefmark{1}DEI, University of Bologna, Via Risorgimento 2, 40136 Bologna, Italy \\
                    \IEEEauthorrefmark{2}Integrated Systems Laboratory, ETHZ, Gloriastr. 35, 8092 Zurich, Switzerland \\
				    \IEEEauthorrefmark{3}SOITEC, Crolles, France\\}}


\twocolumn

\maketitle

\IEEEpeerreviewmaketitle


\begin{abstract}

The performance and reliability of Ultra-Low-Power (ULP) computing platforms are adversely affected by environmental temperature and process variations. 
Mitigating the effect of these phenomena becomes crucial when these devices operate near-threshold, due to the magnification of process variations and to the strong temperature inversion effect that affects advanced technology nodes in low-voltage corners, which causes huge overhead due to margining for timing closure.
Supporting an extended range of reverse and forward body-bias, UTBB FD-SOI technology provides a powerful knob to compensate for such variations.
In this work we propose a methodology to maximize energy efficiency at run-time exploiting body biasing on a ULP platform operating near-threshold. The proposed method relies on on-line performance measurements by means of Process Monitoring Blocks (PMBs) coupled with an on-chip low-power body bias generator. We correlate the measurement performed by the PMBs to the maximum achievable frequency of the system, deriving a predictive model able to estimate it with an error of 9.7\% at 0.7V. To minimize the effect of process variations we propose a calibration procedure that allows to use a PMB model affected by only the temperature-induced error, which reduces the frequency estimation error by 2.4x (from 9.7\% to  4\%). 
We finally propose a controller architecture relying on the derived models to automatically regulate at run-time the body bias voltage. We demonstrate that adjusting the body bias voltage against environmental temperature variations leads up to 2X reduction in the leakage power and a \fix{}{15\%} improvement on the global energy consumption when the system operates at 0.7V and 170MHz.
\end{abstract}

\section{Introduction}
\label{sec:intro}

\IEEEPARstart{P}{ervasive} use of embedded computing platforms in e-Healt, wearables, smart sensors and Internet of Things (IoT) applications is pushing the research community to extensively explore the performance-energy tradeoff. IoT edge computing requires Ultra-Low-Power (ULP) devices consuming few mW and delivering GOPS. 
On the other hand, Moore's law is slowing down, and CMOS scaling does not lead huge energy efficiency benefits any longer\cite{Dreslinski2010a}. This pushes for ever more aggressive voltage reductions, i.e. operating transistors very close to their threshold voltage, an approach known as Near-Threshold Computing (NTC)\cite{Markovic2010}\cite{Rossi2017a}. 
The major challenge that near-threshold IoT devices have to face is operating in many different scenarios that are not always completely predictable at design time. 
More specifically a serious concern is validation on very wide range of operating temperatures. 

This aspect is crucial, as devices implemented in most advanced technological nodes have a strong dependency between environmental temperature, operating frequency and leakage power \cite{Pahlevan2016}\cite{Alioto2017}. This behaviour is caused by a phenomenon called Temperature Effect Inversion (TEI)\cite{Lee2014}. In deep sub-micron technologies the detrimental effect that the temperature has on the maximum frequency of a device is reverted\cite{Pu2010}. Due to TEI \cite{Han2017}, the performance and leakage current have a positive sensitivity to temperature, especially for those devices operating in low-voltage corners. 
Another effect which must be taken into account in ULP devices implemented in recent technological nodes is process variation\footnote{We neglect the effect of aging since self-heating is minimum in ULP devices}.
Accounting for temperature and process variation requires to take very large margins at design time, which causes huge overheads in power due to the needs of buffers for setup and hold time fixing in different corners, often making it impossible to achieve timing closure \cite{Alioto2012a}.

To achieve correct operation and acceptable power-performance ranges in near-threshold CMOS chips, post-fabrication compensation of process and temperature variations is critically needed. Body biasing (BB) is a well-known approach for post-fabrication compensation. It consists in polarizing with a voltage potential the N and P well of CMOS transistors, thereby changing threshold voltage. The performance of a device, in terms of maximum achievable frequency, is directly related with the threshold voltage of transistors, more specifically, when the threshold voltage decreases the maximum frequency of the device increases
\cite{Sundaresan}\cite{Gammie2008}. On the other side, the leakage current increases when the threshold voltage decreases, thus increasing static power consumption.
Body biasing represents an alternative to adaptation of supply voltage, \textcolor{black}{the latter} is less efficient and requires more complex support circuitry, like on chip DC/DC converters or voltage regulators resulting in higher power overheads and coarser granularity\cite{Clerc2015}.

The use of body biasing is twofold: \textit{i)} Forward Body Biasing (FBB) allows to increase the operating frequency of the chip with a minimum power overhead \textit{ii)} Reverse Body Biasing (RBB) could significantly reduce the leakage current when both process and environmental temperature would allow the device to run faster than necessary, given a target application \cite{Rossi2015a}.


The aim of this work is to propose a performance-aware body biasing strategy to compensate for process and temperature variations \textcolor{black}{while guaranteeing with minimum power overhead a user-specified target frequency}. The methodology exploits a calibration procedure which can be executed at boot time, to tune all the involved software models and compensate static performance variations (e.g. process and aging effects). Then, dynamic variations (e.g. temperature) are compensated at run-time by exploiting on-chip frequency measurement obtained from Process Monitoring Boxes (PMB) properly calibrated to minimize the frequency measurement error. Energy efficiency improvement is obtained by exploiting the capabilities of UTBB FD-SOI (Ultra Thin Body and Box Fully Depleted Silicon On Insulator) technology to apply a wide-range forward and reverse body bias voltage. Thanks to on-chip ULP Body Bias Generators (BBG) we modulate the $V_{bb}$ from \SI{-1}{\volt} corresponding to full RBB to $V_{dd}/2$ + \SI{300}{\milli \volt} corresponding to full FBB. The proposed methodology is suitable to be implemented as a software control strategy on any processor featuring both on-chip PMBs and BB generators. The software overhead can be considered negligible, since the controller can be activated with periods in the order of seconds to track temperature variations, and the Body bias regulation requires approximatelye{Rossi2017}. Additionally, exploiting flexibility of software models, higher accuracy can be achieved with respect to pure HW-based controllers\cite{Rossi2017}.


As a preliminary step we operated a complete characterization of the on-chip frequency measurement modules (PMB) available to us. We performed this operation on an advanced chip testing equipment: Advantest SoCV93000 tester system. In this phase we studied the accuracy of the PMBs versus the environmental temperature and we derived a mathematical model correlating their frequency measurement with the actual maximum frequency of the chip. To characterize the PMB in the widest possible temperature and voltage operating range we performed our measurements at three different temperature (T = \{\SI{-20}{\celsius},\SI{25}{\celsius},\SI{80}{\celsius}\}) and supply voltages ($V_{dd}$ = \{\SI{0.5}{\volt},\SI{0.7}{\volt},\SI{0.9}{\volt}\}). At every operating point identified by $V_{DD}$ and T we computed the correlation between the PMB frequency measurement and the real maximum frequency of the device on the entire body bias range (\SI{-1}{\volt} to $V_{dd}/2$ + \SI{300}{\milli \volt}). We repeated this operation on chips belonging to different process corners. 

\textcolor{black}{Note that the scope of the PMB characterization is to establish a relationship between the output of the PMBs and the maximum frequency of the chip. The maximum frequency could deviate from the trend predicted by the PMBs at different supply and body bias voltages. The PMBs itself cannot provide an exhaustive representation of all the reasons that limit the frequency of the device (e.g. critical path going to the memories). However, every mismatch between the output of the PMBs and actual maximum frequency is embedded in the software model constructed during the PMB characterization.}

In a successive phase we developed a strategy exploiting both the PMB and BB models to dynamically modulate the body bias voltage and achieve the desired target frequency against temperature variations. This task is executed by a software body bias controller which probes the performance of the chip and as a consequence programs the body bias generators to apply the required $V_{BB}$. To compensate the model inaccuracy we adopt forward body bias margins that are added to the $V_{BB}$ regulation operated by the controller. These forward body bias margins allow to prevent chip failures when the PMBs are over-estimating the chip performance and the $V_{BB}$ controller is applying less forward body bias than necessary. The model error reduction comes with a cost, adding body bias margins to the $V_{BB}$ regulation increases the leakage current. We demonstrate that with a calibration procedure it is possible to reduce the frequency estimation error by a factor of 2.4X, delivering accurate predictions on chip performance with a minimum leakage cost. We propose a controller architecture based on such models, dynamically adjusting the $V_{BB}$ either to achieve the target frequency or to reduce the leakage current as much as possible. \textcolor{black}{Although the system features two different power domains with body biasing capabilities, to simplify the implementation, the body bias regulation proposed in this paper has been tested and validated on the most computationally capable and power-hungry power domain, that is the cluster. In this scenario only the body bias generator related to the cluster domain is controlled by the regulation loop; the second body bias generator is set to generate a constant VBB = \SI{0}{\volt}.} In this work we demonstrate that the proposed body biasing controlling methodology can achieve up to 2X in leakage reduction and 15\% in energy efficiency improvement at 170MHz and $V_{DD} = 0.7V$ on the related power domain.



The remainder of this work is organized as follows. Section \ref{sec:related} presents an overview of the different methodologies to monitor on-chip the performance and the strategies to dynamically modulate the body bias voltage already presented in the past works.
Section \ref{sec:system} describes the chip used as test vehicle and the most relevant IPs enabling the adaptive body biasing methodology. 
In section \ref{sec:model} we report all the required steps to derive a model for the PMBs and how to reduce its error. 
Section \ref{sec:controller} describes a PMB calibration procedure and the body bias controller; additionally, it also shows two working examples of the dynamic body biasing controller.
Section \ref{sec:results} provides the results in terms of leakage current reduction and energy efficiency improvement when the controller is active. Finally, section \ref{sec:conclusion} contains concluding remarks.

\section{Related Work}
\label{sec:related}

    With the diffusion of Ultra-Low-Power (ULP) devices operating in low voltage corners, robustness to variations introduced by process and temperature has become a critical issue. The problem of variability in this field has been analyzed in the past by several works. The solutions proposed to address this challenge leverage design-time and run-time techniques employed at \textcolor{black}{the} circuit, micro-architectural and architectural levels. These techniques lead to mitigate, improve resiliency, or optimize energy or performance in presence of variations. M. Alioto \cite{Alioto2012a} highlighted the inefficiency to apply conventional design paradigms to ULP-oriented devices, analyzing the impact of process and temperature variation on low power designs operating in the near- and sub-threshold region, and providing guidelines for design of standard cells, memories and microarchitectures.
    Other circuital and architectural solutions to reduce the impact of variability are presented by M. Seok et. al. \cite{Seok2011}, their work mostly focuses on design-time techniques addressing the problem of variations in logic, memories and clock tree.
    
    An orthogonal approach to address the problem of variation is that of compensation. While the aforementioned methodologies rely on decisions that have to be taken at design time (e.g. very conservative supply voltage margins), run-time variation compensation allow to reduce margining, leaving rooms for energy efficiency improvements. 
    Most common approaches to compensate variations at run-time exploit Dynamic Voltage Scaling (DVS) or Dynamic Body Biasing (DBB) as compensation knobs. In many architectures $V_{DD}$ and $V_{BB}$ are controlled in closed loops that rely on circuital parameters tracking (e.g. frequency or temperature). Therefore, probing the system status in terms of maximum achievable frequency and temperature is needed to properly regulate the compensation knob.


\subsection{Probing approaches}


    A powerful approach to estimate system maximum frequency in the context of DVS methodologies is "Razor"\cite{Ernst2003}\cite{Das2005} \cite{Blaauw_2008}. It exploits in-situ error detection on the processor pipeline stages, the functionality of the registers is augmented with "Razor" shadow-latches capable to detect a timing failures and recover from it. D. Bull et. al. \cite{Bull2011} applied Razor to a 32bits microprocessor implemented in UMC 65nm process, reporting 3\% of frequency estimation error.
    Fojtik et al. propose an improved Razor-based solution called "Bubble Razor" \cite{Fojtik2013}, targeting designs operating in low-voltage corners and exploiting flip-flops based datapath to two-phase latches datapath conversion; this allows the use of two-phases clocking methodology, enabling larger timing speculation windows with respect to \cite{Ernst2003}\cite{Das2005}\cite{Blaauw_2008}. The advantage in using these strategies is that they allow \textcolor{black}{to} significantly reduce design margins, mitigating the effect of both global and local process variations when used in combination with compensation approaches. However, if on one hand they enable promising energy savings, on the other hand their application is very intrusive, and it is limited to all those cases where deep knowledge of the architecture is possible, and modification to the RTL are allowed. Very often IPs composing a System on Chip (SoC) are provided as encrypted macros, or simply come as hardened macros from different design teams, hence no access to RTL to modify pipeline stages is possible. 
    

    Other approaches for maximum frequency on-chip probing are based on Critical Path Monitor (CPM). CPM are critical path replicas to which extra delay elements are added to make the path super-critical. Drake et al. \cite{Drake2007} propose a CPM device to track intra-pipeline stage critical path (CP) where 5 different critical path replicas featuring different types of gate (e.g NAND4, NOR3) are synthesized to emulate different sensitivities of datapath logic gates to supply voltage. 
    Tschanz et al. \cite{Tschanz2009} propose a more general approach. Instead of pre-determined logic gate types, configurable buffer delay chains are tuned to emulate largest delay path between pipeline stages, providing a feedback on the circuit maximum frequency. 
    A similar solution is proposed by Clerc et al. \cite{Clerc2015}, to take into account process variability, different CPMs types can be tuned to match the critical path of the device at a given voltage and temperature. Due to the low level of intrusiveness, and since ULP devices are only marginally affected by intra-chip variations (devices in few $mm^{2}$ range) CPM-based approaches can be very effective to estimate the maximum frequency. However, when critical path involves RAM, CP can be rarely emulated by cascading logic gates or with delay buffer chains because of the mixed-signal nature of internal signals, making these methodologies hardly usable in complex designs. 
    Clearly identify a critical path is a difficult task in devices fabricated in deep-sub micrometer technology nodes and operating in low-voltage corners. To improve the accuracy, Beign\'{e} et al. \cite{Mhz2015} propose a more general approach combining the use of \textcolor{black}{Critical Path Replica (CPR)} and timing fault detection, coupling two different performance estimation methodologies. Such frequency tracking approach is implemented in 28nm FDSOI technology node and covers multiple voltage operating points (0.4V to 1.3V). 

    As Shown by Zandrahimi et al. \cite{Zandrahimi2016}, a big improvement in generality of probing approaches is introduced by the adoption of Process Monitoring Blocks coupled with a software model (This solution is adopted in our work). Thanks to hardware sensors based on several types of ring oscillators (e.g. NMOS only and PMOS only), frequency estimation can be performed with an accuracy of 7.6\%. As we demonstrate in this work, by properly calibrating PMB models this error can be further reduced to 4\%. Indeed, models implemented in software can be more flexible than LUT implemented in hardware and \textcolor{black}{better fit} the behaviour of PMBs. 
    Moreover, adopting a PMB-based solution allows also to overcome limitations coming from architecture or CP awareness. Finally, area overhead of this probing methodology is very limited, as visually evident from the example of \figurename{ \ref{fig:micro}}.

\subsection{Actuation approaches}

    Actuation can be seen as the next step of probing, once the performance \textcolor{black}{has} been estimated, some action is performed to align the current behaviour of the system with the desired one.   
    Some approaches dynamically change the operating frequency depending on the specific process/temperature/aging conditions. The solution proposed by Constantin et al. in \cite{Constantin2016} tackles the problem of margins reduction from a different perspective with respect to the adoption of specific circuital solutions \cite{Alioto2012a}\cite{Seok2011}. Timing constraints are relaxed at design-time, simplifying timing closure. Clock frequency is then modulated at run-time \fix{}{according} to the specific critical paths triggered by the executed instructions. 
    Even though this methodology can allow significant energy improvements, its application requires to know both critical paths and delay paths triggered by a certain processor instruction. Moreover, the application of this methodology \fix{}{cannot} ensure a target frequency matching.

    

    Compensation of temperature and process variation has traditionally been performed by supply voltage adaptation. Various strategies exploiting Dynamic Voltage Scaling (DVS) in bulk technologies are presented in \cite{Ernst2003} \cite{Blaauw_2008} \cite{Bull2011} \cite{Fojtik2013}\cite{Bowman2011}, where design-time margins are reduced thanks to the in-situ timing violations detection strategies previously mentioned. Specifically, the maximum performance of the chip is probed and used as feedback to modulate the supply voltage right before the Point-of-First-Failure (PoFF). 
    While DVS can be very effective to reduce design-time margins, compensating run-time performance degradation caused by process and temperature variations, as demonstrated by \cite{Clerc2015}, is a task that can be executed much more efficiently leveraging body biasing.
    
    

    
    
    M. Miyazaki et.al. \cite{Miyazaki} demonstrated the effectiveness of process compensation by exploiting -1.5 RBB to + 0.5 FBB body bias range in a 200 nm CMOS technology. Similarly, Tschanz et. al. \cite{Tschanz2002} implemented in 150nm CMOS technology node an adaptive body biasing scheme (-0.5V to + 0.5V) for process compensation. More recently, DBB has been also used to compensate not only for process and aging variation but also to compensate short-term performance variation caused by temperature changes.
    Kumar et.al. \cite{Kumar2008a} developed an algorithm for temperature compensation in the range 35\textdegree C to 65\textdegree C, exploiting body biasing in devices fabricated in 65nm and 45nm technology nodes. Similarly, Tschanz et. al. \cite{Tschanz2007} used body biasing for temperature compensation of a TCP/IP processor in 90nm technology operating at 1V in a range from 60\textdegree C to 80\textdegree C.
    More recently, Kang et al. \cite{Kang2010} propose a PLL-based performance feedback circuital solution exploiting body biasing, implemented in IBM 130nm to compensate process, temperature and aging induced variations.
    In successive works, Kumar et al. \cite{Kumar2011} and Ono et al. \cite{Ono} presented two hybrid methodologies exploiting adaptive forward body biasing combined with DVS to maintain optimal performance of a device against variations introduced by aging. 
     More recently, Gammie et. al. \cite{Gammie2008} exploited a -0.5V to 0.5V body biasing for process variation compensation and for high performance and low-power states enabling. 
     
    All the listed approaches refer to performance compensation for devices implemented in bulk technologies. Unfortunately, such technologies only achieve good results either in limited temperature ranges or leveraging too small body bias ranges. 
    In deep-submicron bulk technologies, maximum body bias range is limited by p-n junction leakage and potential latch up. In FinFET technologies \cite{Rossi2015a}, instead, the lack of an easy way to access the back gates represents the main obstacle to body bias voltage application \cite{Rossi2015a}. In near-threshold, the impact of temperature variations is huge and cannot be fully compensated with the limited body biasing capabilities provided by the bulk CMOS technologies.
    On the contrary, 28nm UTBB-FDSOI provides a very powerful knob for process and temperature variations compensation because of its wide-range body biasing capability (theoretically from \SI{-3}{\volt} to \SI{3}{\volt}). A good demonstration of the body biasing capabilities in FDSOI technology is provided by Clerck et al. \cite{Clerc2015}, however the proposed design is fabricated with LVT cells, allowing only forward body biasing of the transistors in the range \SI{0}{\volt} to \SI{3}{\volt}. Contrarily, as anticipated by the analysis reported by Rossi et al. in \cite{Rossi2017}, and as implemented in this work, a processor implemented in UTBB FDSOI with RVT can exploit wide-range body biasing to implement advanced power management and compensation strategies.

   In this work we fully characterize the PMB sensors, studying the correlation with the maximum frequency of the device in presence of process variations (i.e. among multiple chips belonging to different process corners), and demonstrating that with PMB sensors and properly calibrated software model it is possible to obtain a performance estimation accuracy comparable with the probing approaches previously described (4\% at 0.7V, including temperature variations), with a negligible overhead on the device area, and with an extreme low level of intrusiveness. We then present a body bias controller capable to boost the performance through FBB and enter low leakage full-state-retention modes through RBB. Thanks to a finely calibrated control strategy implemented in software, we demonstrate how it is possible to compensate dynamic performance degradation and reduce the leakage current on an a real embedded platform, exploiting the higher efficiency of compensating temperature and process variations with body biasing\cite{Rossi2017}\cite{Clerc2015}.

\section{PULP System}
\label{sec:system}


\subsection{UTBB FD-SOI technology}

PULPv3 has been \fix{}{implemented} in 28nm Ultra-Thin Body and Box Fully Depleted SOI technology (UTBB FD-SOI) from STMicroelectronics. This technology features an improved channel electrostatic control thanks to thin-film technology, reducing leakage currents and Short Channel Effects (SCE). On this technology, for the same leakage current target the threshold voltage can be strongly scaled, thanks to the ultra-thin buried oxide. This ensures low variability when circuits operate close \fix{}{to the} threshold voltage of transistors. Ultra-thin buried oxide also enables the use of very wide body biasing range from -3V to 3V using conventional and flip well transistor\cite{Rossi2015a}\cite{Flatresse2013}. In UTBB FD-SOI technology channel length modulation (i.e. poly biasing) is used at design time to statically optimize circuit critical path. In the case of PULPv3 implementation, we used the conventional-well flavor of the technology
\cite{Rossi2015a} and the entire system is implemented with the same type of Poly-Biasing 0 (PB0) Regular Voltage Threshold (RVT) cells.

\subsection{Architecture}

\begin{figure}[t]
    \centering
    \includegraphics[width = \columnwidth]{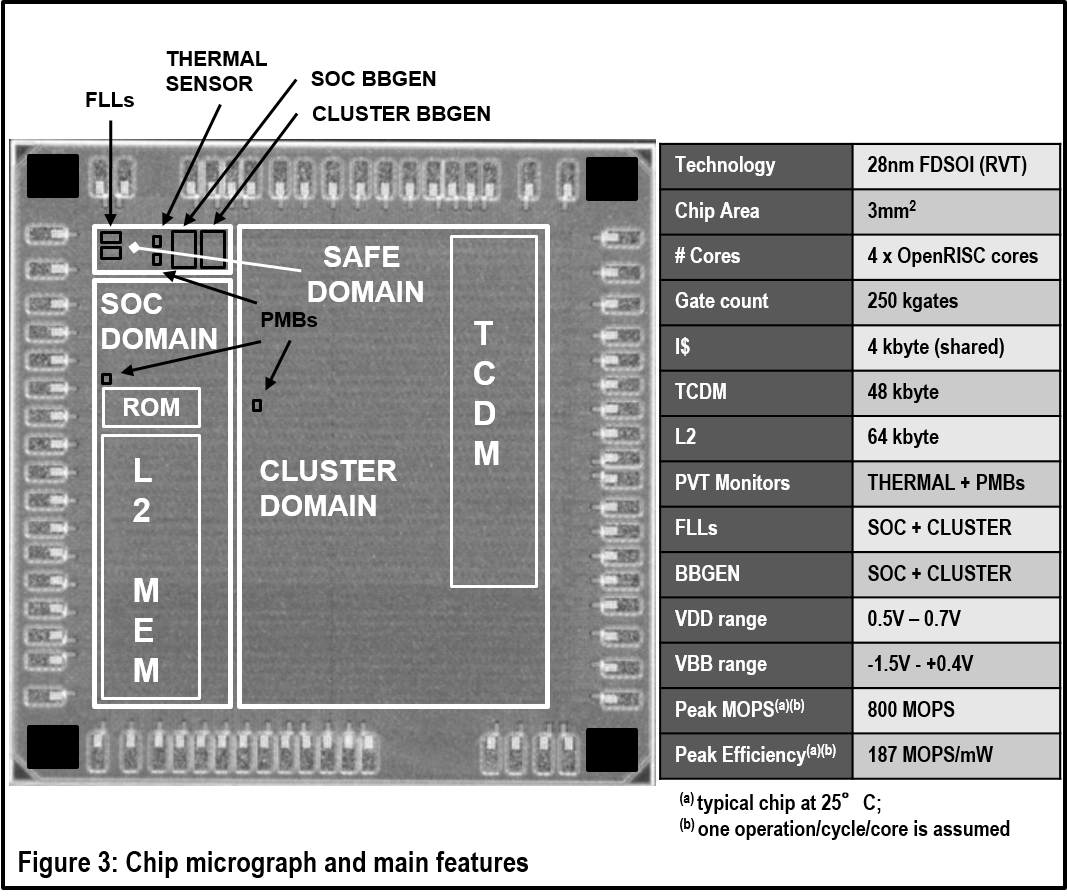}
    \caption{This figure shows a micrograph of the chip where it is possible to identify: the SoC domain, the cores cluster and the safe domain. Additionally, it is possible to note the PMB sensors in every power domain and PMB controller. Table on the right shows the main architectural  features of the PULPv3 chip.}
    \label{fig:micro}
\end{figure}

\begin{figure}[tb]
    \centering
    \includegraphics[width = \columnwidth]{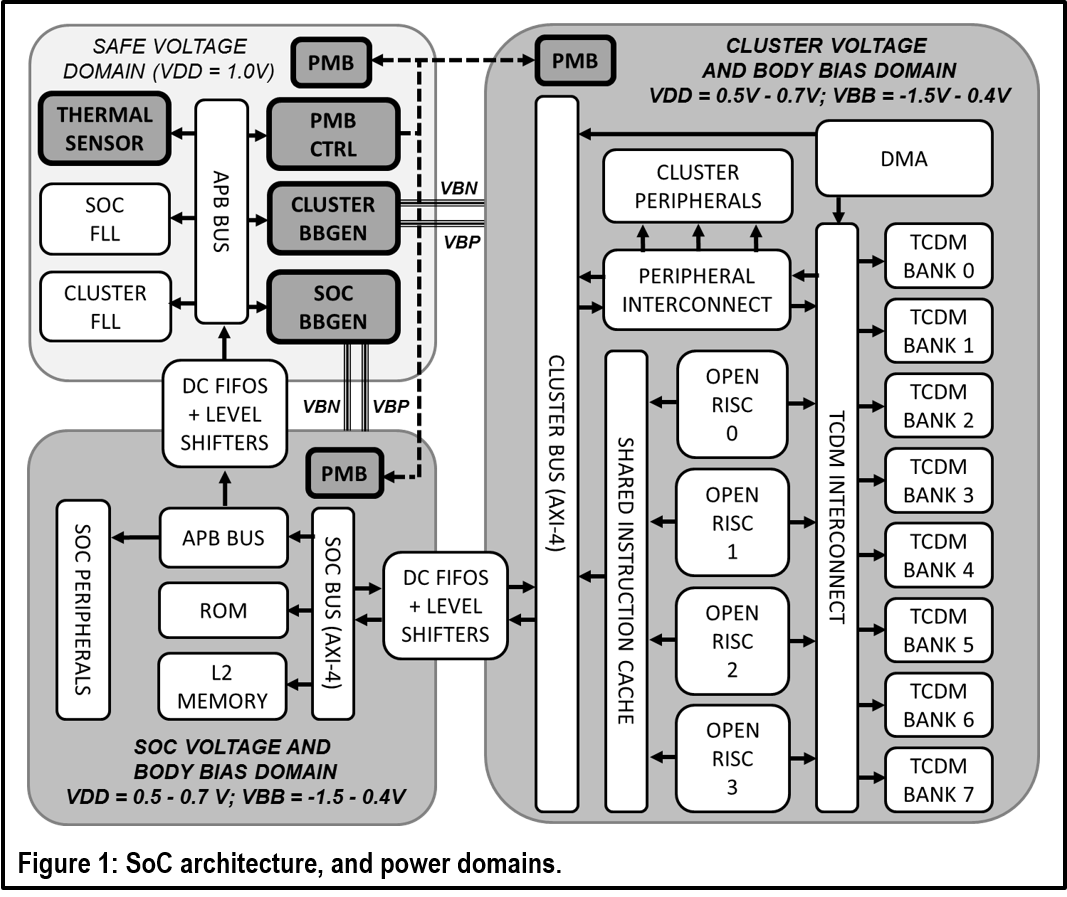}
    \caption{Architecture of PULPv3 system. The picture shows the three different power domains: the SoC domain, the cores cluster and the safe domain}
    \label{fig:archi}
\end{figure}

Parallel Ultra-Low-Power platform \cite{Conti2015} version 3 (PULPv3) \cite{Rossi2017} is a multi-core SoC for ULP applications operating in near-threshold to achieve extreme energy efficiency on a wide range of operating points. The SoC is built around a cluster featuring four cores and \SI{64}{\kilo byte} of L2 memory. The cores are based on a highly power optimized micro-architecture implementing the OpenRISC-32bit ISA featuring 4kB of shared instruction cache. The cores do not have private data caches, avoiding memory coherency overhead and increasing area efficiency, while they share a L1 multi-banked Tightly Coupled Data Memory (TCDM) acting as a shared data scratchpad memory. The TCDM features 8 4kB SRAM  banks and 8 1kB \fix{}{Standard Cell Memory (SCM)} banks connected to the processors through a single clock latency non-blocking interconnect, implementing a word-level interleaved scheme to minimize banking conflict probability. Off-cluster (L2) memory latency is managed by a tightly coupled DMA featuring private per-core programming channels, ultra-low programming latency and lightweight architecture optimized for low-power and high transfer efficiency. \figurename{\ref{fig:archi}} shows the architecture of the system while
\figurename{ \ref{fig:micro}} shows a micrograph of the chip.

Three isolated power domains enable advanced power management: \textit{i)} The "Safe Voltage Domain") hosting the Frequency Locked Loop generators, the two Body-Bias Generators for the SoC and Cluster regions, the PMB Controller and additional infrastructural control logic \textit{ii)} The  "SoC Body-Bias Domain" \textit{iii)} The "Cluster Body-Bias Domain". Each domain is monitored by a PMB, which will be described in the next section. In our tests, we will focus only on the Cluster Domain, applying $V_{bb}$ = \SI{0}{\volt} to the other body-bias domains.

\subsubsection{Process Monitor Boxes}

A Process Monitor Box (PMB) is as an on-chip sensor based on ring oscillators, connected to the system through a memory mapped interface. It provides a measurement of the maximum frequency achievable by the device. These sensors are designed and optimized to behave consistently with the other library logic gates, emulating also performance variations induced by temperature and process variations. 
Since PULPv3 has been implemented with PB0 cells, the ring oscillators of the PMB sensors instantiated in each isolated power domain are implemented with the same type of PB0 cells. 

\subsubsection{Body-Bias generators}

The body bias voltage is modulated thanks to a fully integrated body-bias generator. Such body bias generator is capable to \textcolor{black}{supply an area of \SI{1}{\milli\metre\squared}.} It applies a fine-grain body-bias voltage with a minimum step of \SI{50}{\milli \volt} and can cover a voltage range from \SI{-1.5}{\volt} to $V_{DD}/2$ + \SI{300}{\milli \volt} in a 4.15$\mu W$ power budget. This device has been designed to target ULP system on chips\cite{Blagojevic2016} and further optimized to maximize the energy efficiency\cite{Rossi2017}. \textcolor{black}{ For positive regulation of both wells it uses a push-pull approach. Negative regulation on the p-well is obtained by means of a dual phase charge pump. Two control loops allow to monitor the impact of the leakage and compensate for wells discharge. The comparators of the feedback use resistive Digital to Analog Converters (DAC) as references. To reduce the power consumption of the BB gen to \SI{4.5}{\micro \watt}, its operation is duty-cycled, entering a sleep mode where only the leakage is monitored.} The generator is programmed through a series of memory mapped registers and can independently bias with different voltages the Nwell and the Pwell of the transistor. Table \ref{tab:BBgen} reports the main features of the body bias generator.

\begin{table}[h]
    \centering
    \begin{tabular}{l|c}
        \toprule
        BBGEN Area & 0.00913 $mm^2$ \\
        
        BBGEN Supply & 1.8V $V_{DDIO}$,  1V $V_{DD}$    \\
       
        Power & 4.15 $\mu W$ \\
        
        Transition time N-WELL & 23$\mu s$ \\
        Transition time P-WELL & 11.5$\mu s$ \\
        
        Transition energy & $\leq$25nJ\\
        \bottomrule
    \end{tabular}
    \\[5pt]
    \caption{Main features of the body bias generator.}
    \label{tab:BBgen}
\end{table}

\subsection{PULPv3 Embedded Test System}

\begin{figure}[t]
    \centering
    \includegraphics[width = \columnwidth]{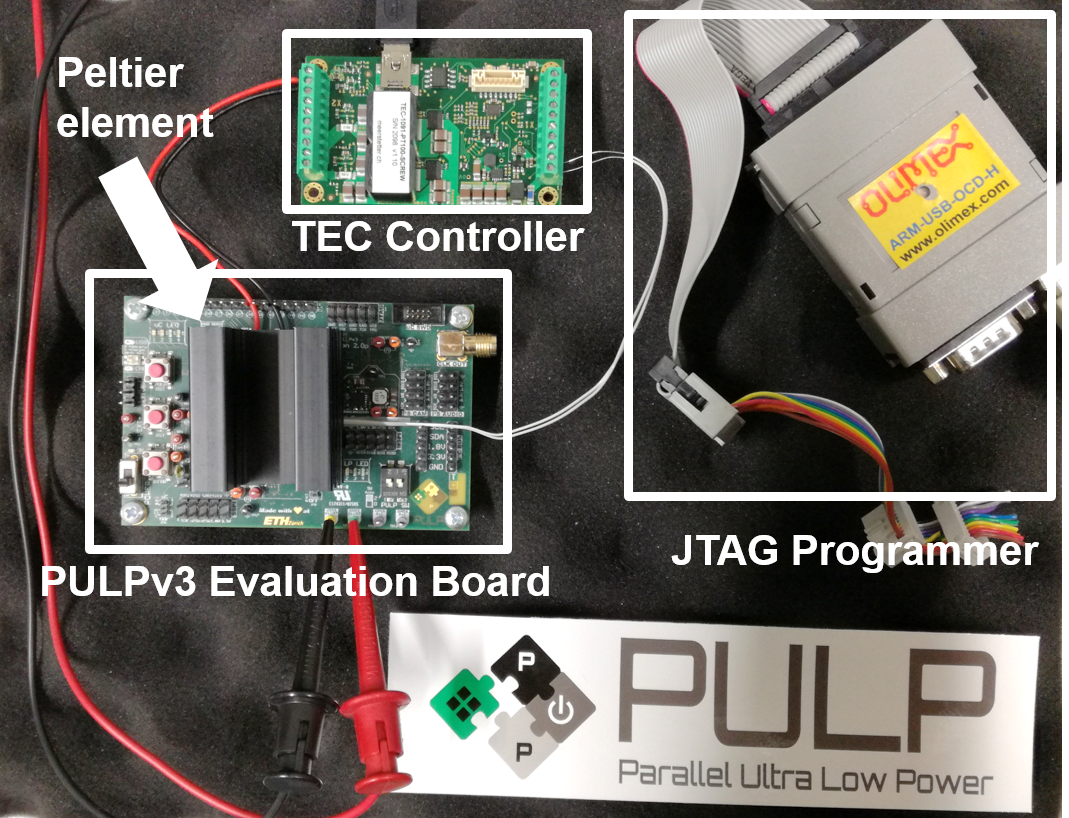}
    \caption{PULPv3 body-bias controller testing setup. In the picture it can be observed the PULPv3 evaluation board hosting the PULPv3 chip and the JTAG programmer. Picture also shows the Peltier's element cell and the controller used to enforce a known temperature on the chip package}
    \label{fig:embedded_setup}
\end{figure}

Since the final goal of this work is to develop a fully automated body-bias control strategy, the analysis of the PMBs has been performed on the chip testing equipment while the proposed methodology has been validated on the PULPv3 evaluation board; this is an embedded platform that can be used to develop and test software applications like commercial microcontroller evaluation boards. The body bias voltage and the leakage current have been measured from the PULPv3 board jumpers by means of a power analyzer. The external temperature variation have been simulated by enforcing a temperature on the chip package by means of a Peltier's element cell regulated by an industrial TEC controller. The Peltier's cell can modulate the chip package temperature in a range \SI{10}{\celsius} to \SI{80}{\celsius}; the complete testing setup is reported in \figurename{ \ref{fig:embedded_setup}}.

\section{Model}
\label{sec:model}

\subsection{Experimental Setup}
\label{sec:setup}


The PMB models derivation, as well as the body bias model described in the following sections are performed with an Advantest SoCV93000 tester system, in connection with a Thermonics 2500E temperature forcing system, able to force an environment temperature ranging from \SI{-80}{\celsius} to \SI{220}{\celsius}. 
Since we are interested in using the body-biasing voltage as independent variable both for temperature and process variations compensation, we structured the measurements as follows: \textit{i)} We defined the operating point (OP) in terms of \{Supply voltage $V_{dd}$ , Temperature\}. More specifically, the voltage corners are: $V_{dd}$ = \{\SI{0.5}{\volt},\SI{0.7}{\volt},\SI{0.9}{\volt}\}. Temperature corners are: T = \{\SI{-20}{\celsius},\SI{25}{\celsius},\SI{80}{\celsius}\}; 
\textit{ii)} For each OP we swept the body-biasing voltage ($V_{bb}$) in the range \SI{-1}{\volt} to $V_{dd}/2$ + \SI{300}{\milli \volt} using the body-bias generator.
At every operating point we measured: the leakage power ($P_{LKG}$), the active dynamic power ($P_{DYN}$), the total power ($P_{TOT}$) and maximum frequency ($F_{MAX}$) achievable by the device.
We measured the active dynamic power ($P_{DYN}$) as the difference between the total power ($P_{TOT}$) and the leakage power ($P_{LKG}$).
The total power ($P_{TOT}$) has been measured as the power consumption when the device is executing a benchmark application and the leakage power ($P_{LKG}$) as the power consumption when the device is not clocked. 

We extracted the maximum operating frequency ($F_{MAX}$) by means of a carefully crafted benchmark (i.e. a sequence of arithmetic operations and memory stores), able to trigger the most critical paths\footnote{The critical path has been identified by the timing analysis in the communication between the cores and the scm memory} of the circuit. We verified that the result of the benchmark was returned with the correct timing, and the End-Of-Computation\footnote{Physical output pin of the device which certifies that the system completed all the operations and properly entered a known final state.} signal was properly asserted. 
As cross-check, we verified that the result of the benchmark returned a valid check-sum.

\subsection{PMB characterization}
\label{analysis}


First part of the analysis focuses on the characterization of the performance monitoring blocks placed in the power domain of interest, that is the one hosting the core cluster. The aim of this step is to obtain a model allowing to correlate the response of the frequency probes (PMBs) to the maximum frequency of the device. To obtain the model, we simultaneously measured both the response of the PMBs ($F_{PMB}$) and the maximum chip frequency ($F_{MAX}$) versus the full body bias range. This operation has been repeated per each operating point ($V_{DD}$, $T$). \tablename{ \ref{tab:meas_details}} reports more details regarding the measurement operating conditions.

\begin{table}[htb]
    \centering
    \begin{tabular}{c|c|c|c}
    \toprule
        PULP $V_{DD}$ & $V_{BB}$ & $T_{op}$ & Process corner\\ \midrule
        \SI{0.7}{\volt} & \SI{-1}{\volt} to \SI{650}{\milli \volt} & \SI{25}{\celsius} & Typical \\
    \bottomrule
    \end{tabular}
    \\[5pt]
    \caption{Measurement operating conditions under which the PMB analysis has been performed}
    \label{tab:meas_details}
\end{table}

\begin{figure}[tb]
    \centering
    \includegraphics[width = \columnwidth]{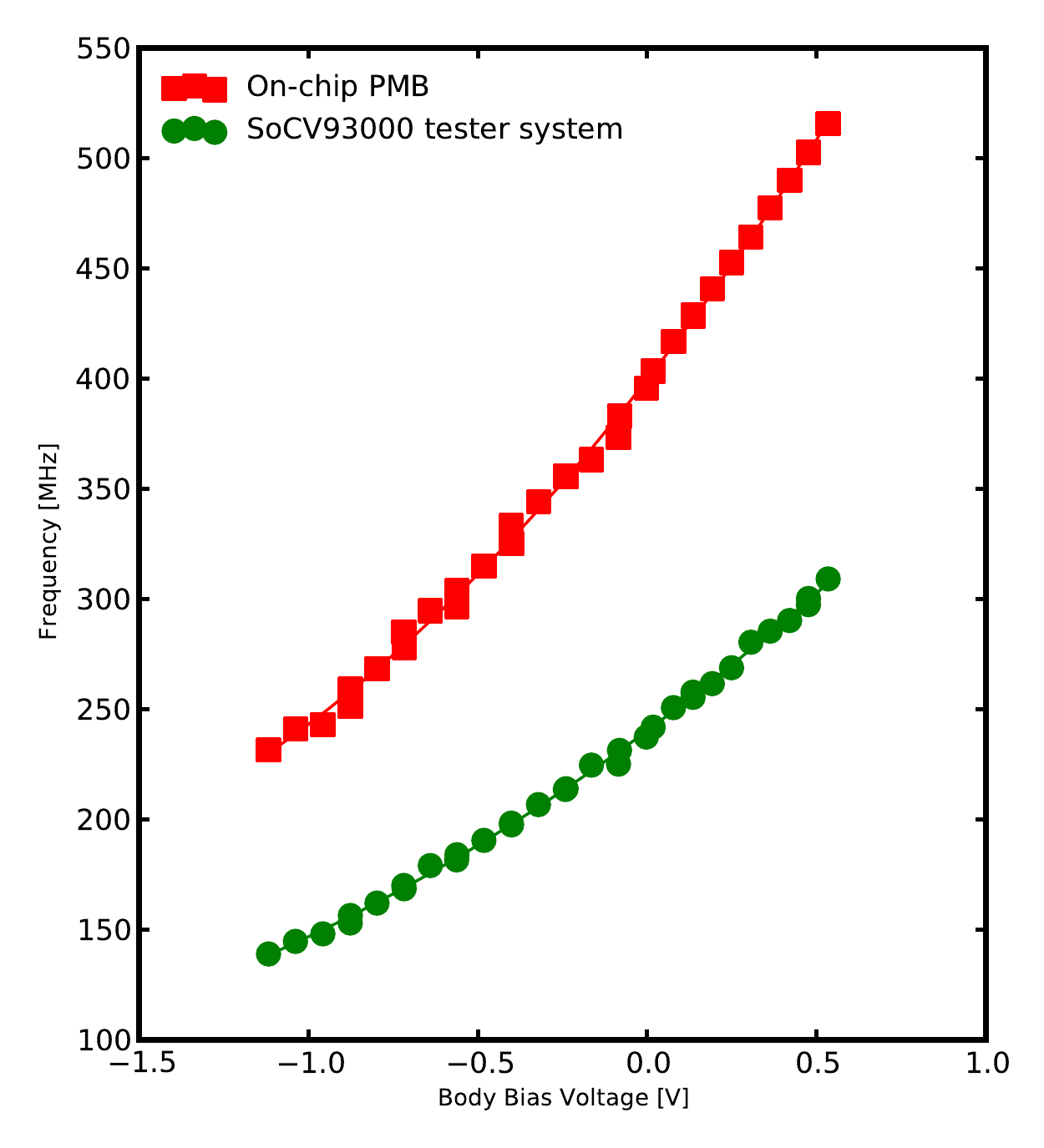}
    \caption{This plot compares the maximum achievable frequency and the PMB frequency estimation for a "Typical" PULPv3 chip at \SI{0.7}{\volt} and \SI{25}{\celsius}.}
    \label{fig:frequency}
\end{figure}

The aim of this step is to obtain an evaluation of the PMB accuracy. To perform the measurement in a well-defined environment, the temperature is externally enforced and kept constant by the temperature forcing system. The measurements that we performed in this phase are referred to a single chip, this allows to exclude a priori the process variations as possible source of error. Since the entire set of PULPv3 chips have been pre-characterized in terms of maximum performance, we know if the device in exam belongs to a \textit{Fast}, \textit{Typical} or \textit{Slow} process corner. \figurename{ \ref{fig:frequency}} reports the measurement of the $F_{PMB}$ and $F_{MAX}$ versus $V_{BB}$ for a typical device supplied at \SI{0.7}{\volt}.

\begin{figure}[tb]
    \centering
    \includegraphics[width = \columnwidth]{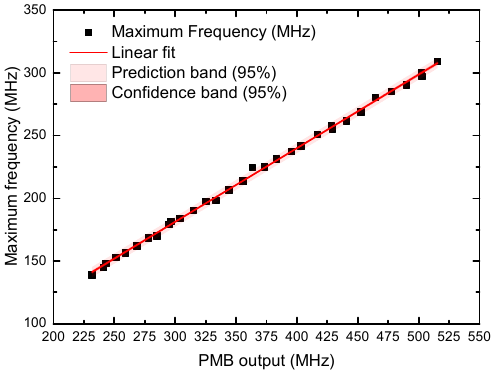}
    \caption{This plot shows the correlation between the maximum frequency and the PMB frequency estimation for a typical chip at \SI{0.7}{\volt} and \SI{25}{\celsius}. The data have been fitted with a linear model, the parameters of the fit are: y=mx+q where m=0.59 and q=5.19, R-Square (COD) = 0.998.}
    \label{fig:correlation}
\end{figure}

In the second phase of the PMB characterization we evaluated the correlation between $F_{PMB}$ and $F_{MAX}$. As shown in  \figurename{ \ref{fig:correlation}}, the model we found is the linear function reported in \ref{eq:model}. In \figurename{ \ref{fig:correlation}} is reported the linear fit of the data.

\begin{equation}
\label{eq:model}
    F_{MAX} = C_{corr} F_{PMB} + F_{0}
\end{equation}


We repeted the same measurements in different operating corners proving the robustness of the model: \tablename{ \ref{tab:model}} reports the parameters of the model.

\begin{table}[hbt]
    \centering
    \begin{tabular}{l|c|c|c}
    \toprule
        $V_{dd}$ & \SI{0.9}{\volt} & \SI{0.7}{\volt} & \SI{0.5}{\volt}\\ \midrule
        $C_{corr}$ & 0.6 & 0.59 & 0.47 \\
        $F_{0}$ & 8.72 & 5.19 & 3.21\\
        R-Square & 0.995& 0.998 & 0.998\\
        \bottomrule
    \end{tabular}
    \\[5pt]
    \caption{Parameters of the model for a typical chip at three different supply voltages and \SI{25}{\celsius}.}
    \label{tab:model}
\end{table}

As a final step for this preliminary PMB characterization we estimated the error of the model, which is represented by the residuals of the measurements with respect to the fitting curve. Table \ref{tab:temp-proc} (\textit{Process-aware Temperature-aware} model) shows the maximum errors we report for the correlation model, for a single device. Comparable errors are reported also in \cite{Beigne2015}, where a similar study performed on a DSP architecture implemented with the same FD-SOI technology, exploiting a performance monitor system based on Timing Fault Sensors (TMFLT).


\subsection{Temperature Variations}
\label{temperature}


Once the robustness of the PMB model has been proved in a given operating condition, it is possible to generalize the analysis to cover a wide temperature operating range. The measurements described in this section are performed on a single chip, supplied with a given voltage ($V_{DD}$), at multiple temperatures. The approach is the same we followed in the previous case, the only difference is that the fitted data are now related to three different temperatures, specifically T = \{\SI{-20}{\celsius},\SI{25}{\celsius},\SI{80}{\celsius}\}.

\begin{table}[h]
\centering
    \begin{tabular}{l|c}
    \toprule
        $V_{dd}$ & \SI{0.7}{\volt}\\ \midrule
        $C_{corr}$ & 0.59 \\ 
        $F_{0}$ & 5.42 \\ 
        R-Square & 0.996\\
    \bottomrule
    \end{tabular}
    \\[5pt]
    \caption{Parameters of the model fitting the data for a single chip, at a given supply voltage, at three different temperatures, T = \{\SI{-20}{\celsius},\SI{25}{\celsius},\SI{80}{\celsius}\}.}
    \label{tab:model_temp}
\end{table}

\begin{figure}[t]
    \centering
    \includegraphics[width = \columnwidth]{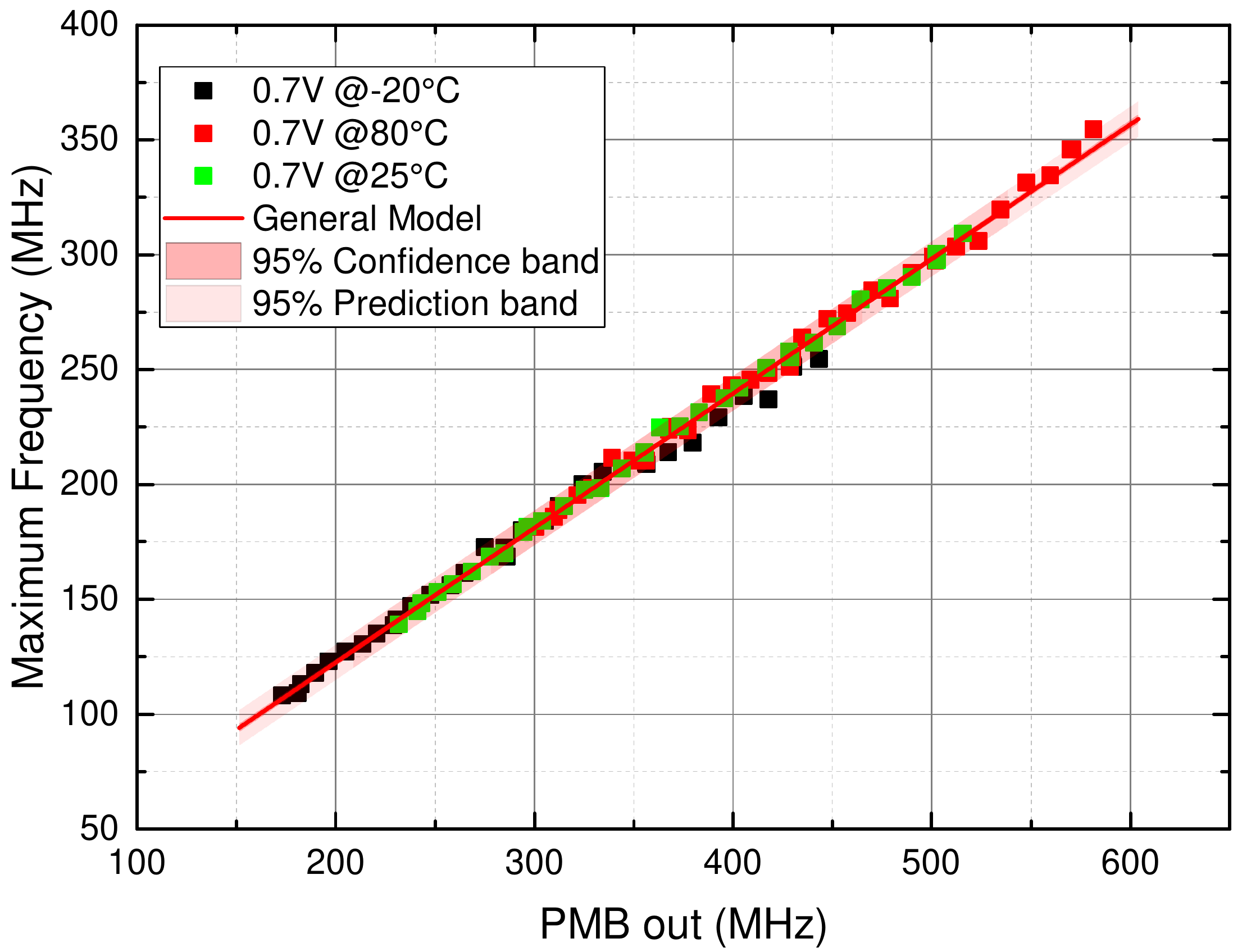}
    \caption{In this plot are shown all the measurements performed at a single supply voltage $V_{dd}$ = \SI{0.7}{\volt} and three different temperatures T = \{\SI{-20}{\celsius},\SI{25}{\celsius},\SI{80}{\celsius}\}. The red solid line represents the curve  fitting the data measured at the three different temperature.}
    \label{fig:perchip_model}
\end{figure}

\begin{figure}[t]
    \centering
    \includegraphics[width = \columnwidth]{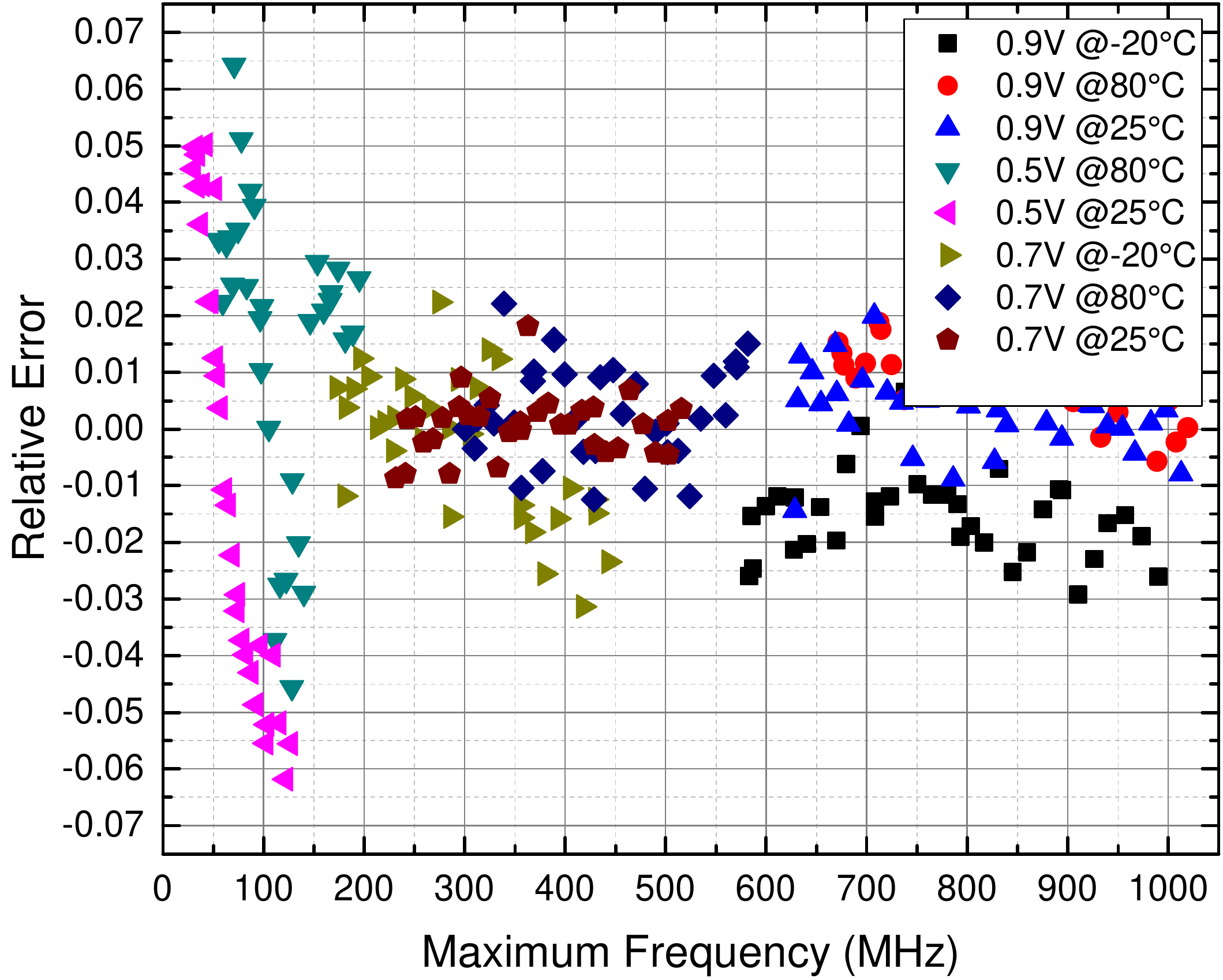}
    \caption{This plot shows the distribution of the relative error versus the maximum frequency when the general model is used to convert the PMB output value in $F_{max}$.}
    \label{fig:perchip_error}
\end{figure}

\figurename{ \ref{fig:perchip_model}} shows the global linear fit. The red solid line, which is described by the equation \ref{eq:model}, represents the general model fitting the data. Table \ref{tab:model_temp} reports the parameters of the model in this operating condition, as well as the R-Square. As in the previous case, we evaluated the error. \figurename{ \ref{fig:perchip_error}} shows the residuals with respect to the fitting curve. As expected, a model which fits data belonging to measurements at various temperatures is affected by a larger error. This phenomenon is caused by the fact that we are assuming that the maximum frequency is limited only by the PMB related to the PMOS transistors, which is usually slower than PMB related to the NMOS ones. Correlation between the maximum frequency of the device and a linear combination of both PMB could improve the accuracy.

\subsection{Process variations}
\label{process}

\begin{figure}[b]
    \centering
    \includegraphics[width = \columnwidth]{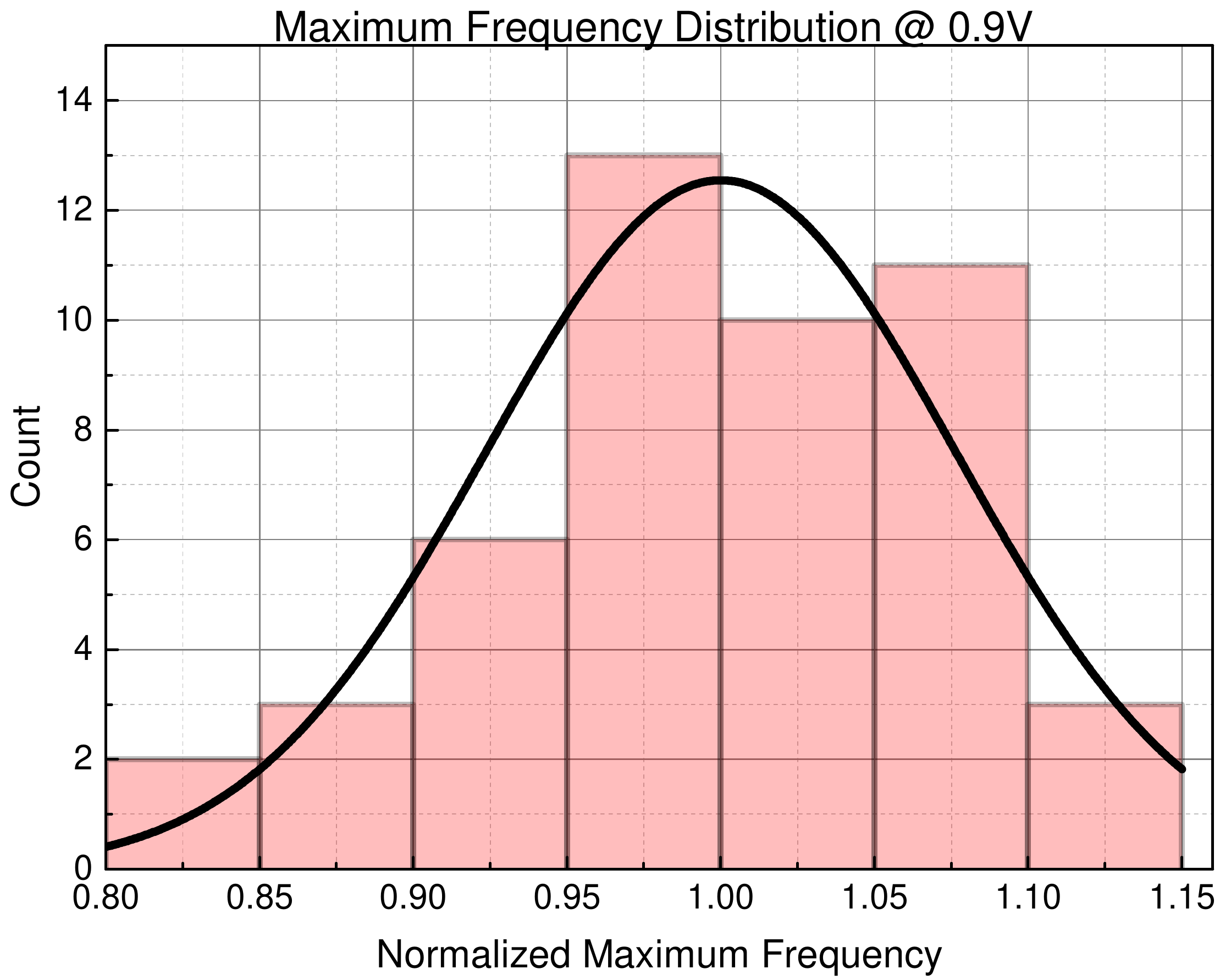}
    \caption{Normalized distribution of the maximum frequency for the entire population of 48 chips at \SI{0.9}{\volt}, \SI{25}{\celsius}.}
    \label{fig:frequency_distribution}
\end{figure}

Operating temperature is not the only factor affecting device performance operating in near threshold. In this section we will present a short discussion on the performance variance among different chips fabricate in the UTBB FD-SOI technological node caused by different process corners. Then, we will show how the performance monitoring methodology described so far can be further generalized to cover also multiple chips belonging to various process corners. Particular attention should be given to the nature of the performance variation related to the technological process, the variations introduced by the process can be classified in two types: \textit{i)} The Inter-chip variations, that can be observed in terms of performance gaps between different devices, as shown in \figurename{ \ref{fig:frequency_distribution}} \textit{ii)} Intra-chip variations, as demonstrated by \cite{Zandrahimi2016}, resulting in different first N critical path, that can affect the consistency between the behaviour of the circuit and an on-chip performance monitor, confirmed also by our analysis.


\begin{figure}[t]
    \centering
    \includegraphics[width = \columnwidth]{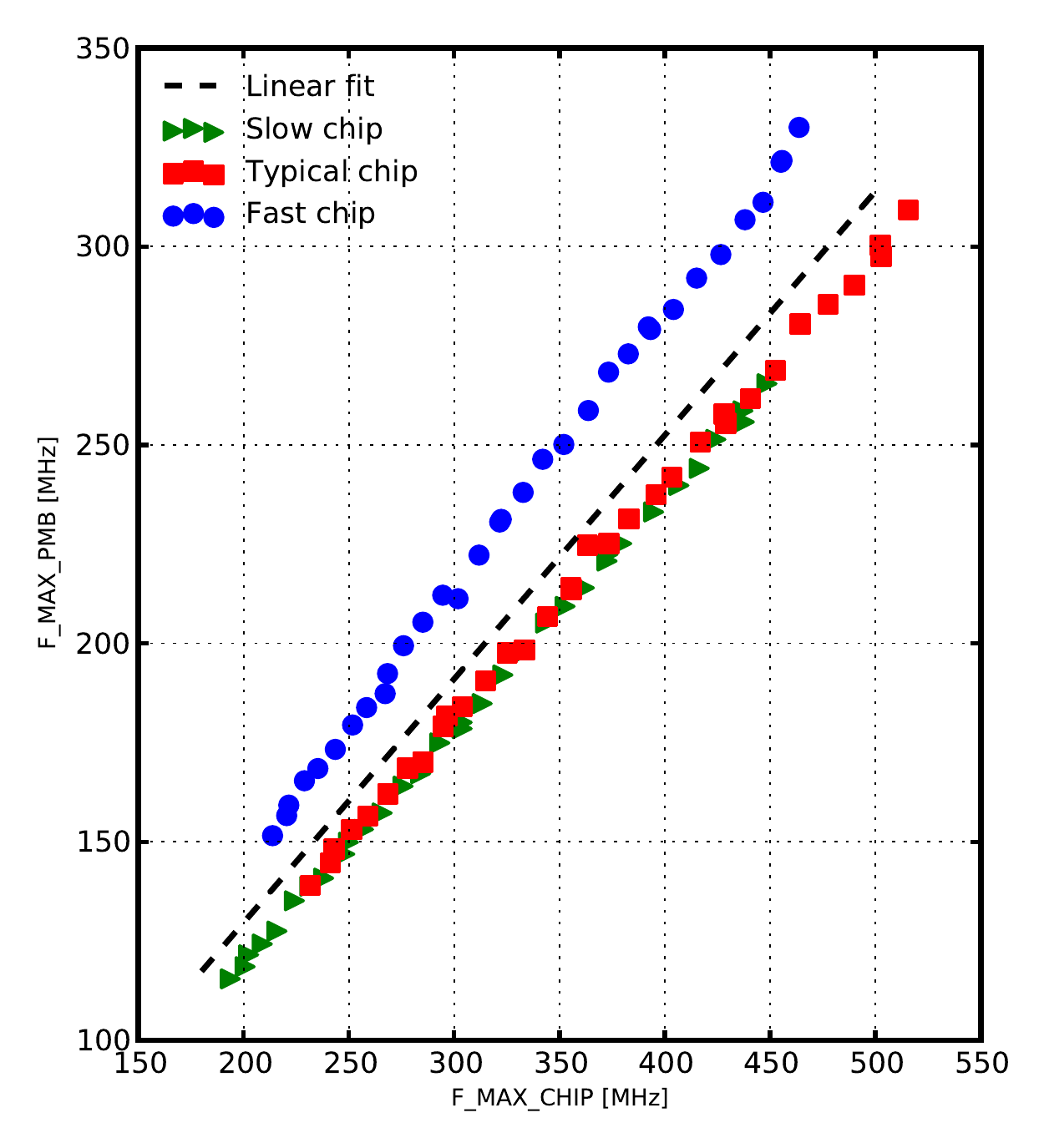}
    \caption{In this plot are represented the data of 3 devices belonging to \textit{Fast}, \textit{Typical} and \textit{Slow} process corners, the dashed line represents the model fitting the data-set}
    \label{fig:process_model}
\end{figure}

As in the case of one chip at multiple temperatures, the model has been obtained by performing a global fit on the data related to chips belonging to different process corners, finding the best function describing them with minimum error. \figurename{ \ref{fig:process_model}} shows the PMB analysis related the chips in exams and the fitting curve.

It can be noted that also in this case the model is represented by the same linear function \ref{eq:model} shown previously. However, in this case the fitted data have a larger variability, and as a consequence, we observed a larger error and lower R-square. Table \ref{tab:model_temp_process} shows the values of the parameters and the R-Square.
When the chip was fabricated the technology node was at a very early stage, according to \cite{Mackay2006}, the variance introduced by the manufacturing process is expected to decrease with the maturity of the process.


\begin{table}[htb]
\centering
    \begin{tabular}{l|c}
        \toprule
        $V_{dd}$ & \SI{0.7}{\volt}\\ \midrule
        $C_{corr}$ & 0.614 \\
        $F_{0}$ & 6.86 \\
        R-Square & 0.88\\
        \bottomrule
    \end{tabular}
    \caption{Parameters of the model fitting the data related to chips belonging to different process corners, at a given supply voltage, at three different temperatures.}
    \label{tab:model_temp_process}
\end{table}

\begin{figure}[htb]
    \centering
    \includegraphics[width = \columnwidth]{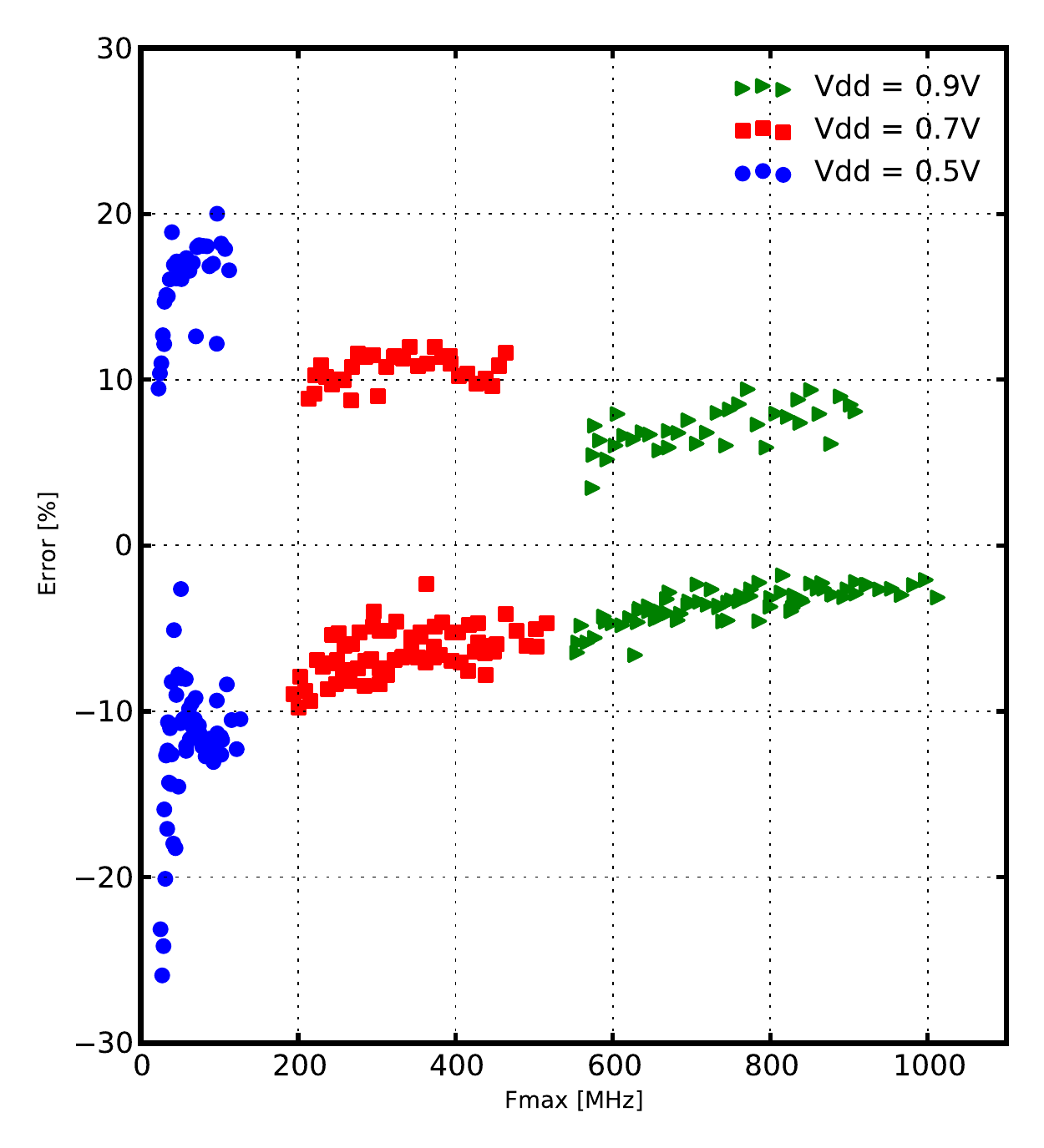}
    \caption{This plot shows the distribution of the relative error versus the maximum frequency when the process-independent model is used to convert the PMB output value.}
    \label{fig:process_error}
\end{figure}



\subsection{Body-Bias Model}
The final goal of this study is to develop a performance-aware body bias regulation system. To this end, it is necessary to derive a model which links a variation of the $V_{BB}$ to the $F_{max}$ variation. Once the relationship between these two parameters has been obtained, given a performance gap to be compensate, it is possible to determine the necessary amount of body bias voltage to do it.

\begin{figure}[htb]
    \centering
    \includegraphics[width = \columnwidth]{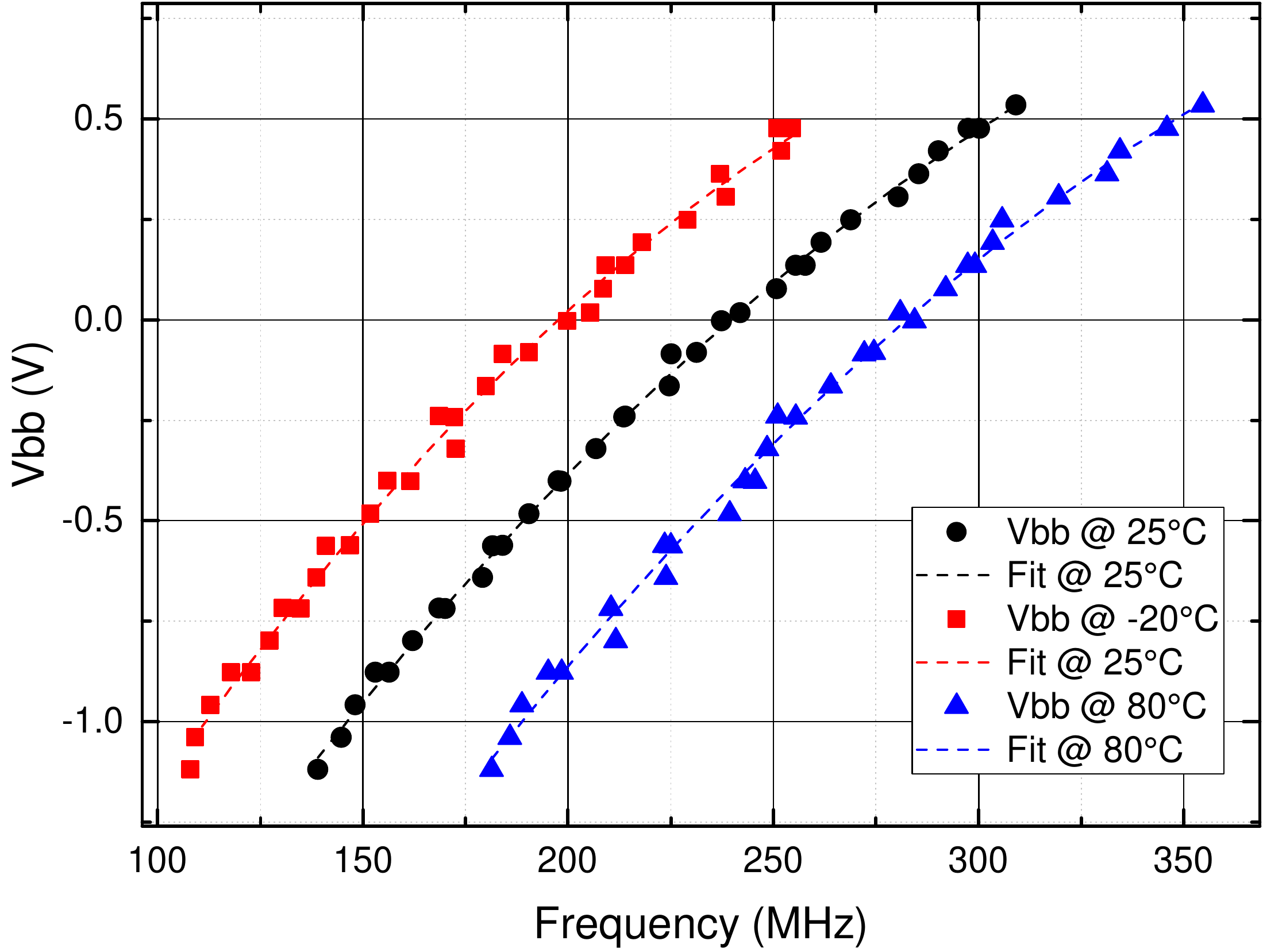}
    \caption{This plot shows the relation between the maximum frequency of a typical device and the body-bias voltage at three operating points: Vdd = \SI{0.7}{\volt} and T = \{\SI{-20}{\celsius},\SI{25}{\celsius},\SI{80}{\celsius}\}.}
    \label{fig:freqvbb}
\end{figure}

It is very important to note that the use of body-bias for this application is twofold: \textit{i)} in an ideal case, that is assuming no errors on the $F_{max}$ estimation, $V_{BB}$ represents the knob to simply change the performance of the device \textit{ii)} in a context where $F_{max}$ is affected by uncertainty because of the reasons previously described, $V_{BB}$ margins can be used to compensate $F_{max}$ negative residuals; in other words, all those conditions where the $F_{max}$ is overestimated by the model. \figurename{\ref{fig:freqvbb}} shows the relationship between $F_{max}$ and $V_{BB}$ at different temperatures.

\begin{figure}[tb]
    \centering
    \includegraphics[width = \columnwidth]{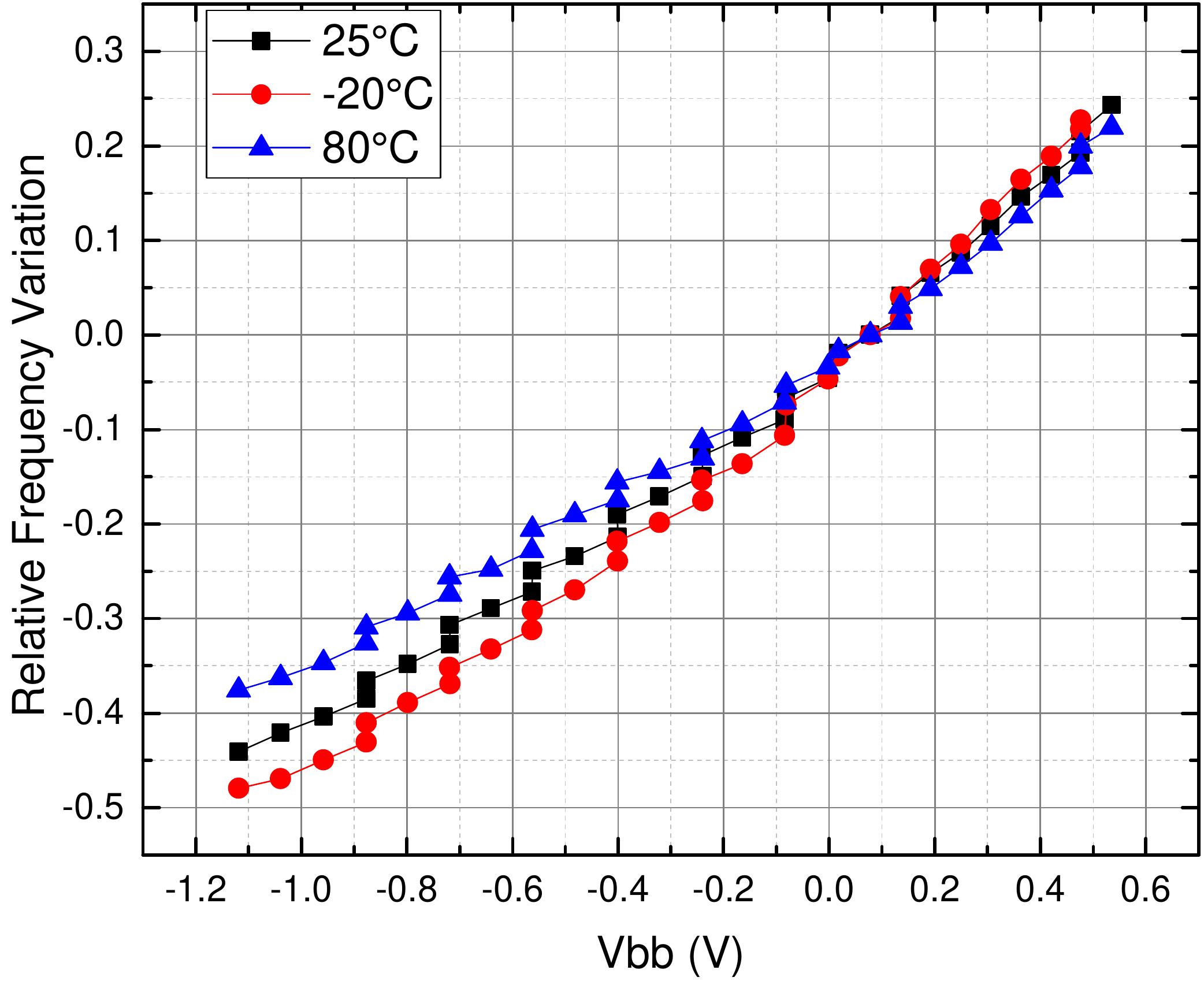}
    \caption{In this plot is represented the relative frequency variation versus the body-bias voltage, with respect to the $V_{bb}$ = \SI{0}{\volt} condition. The operating points are: Vdd = \SI{0.7}{\volt} and T = \{\SI{-20}{\celsius},\SI{25}{\celsius},\SI{80}{\celsius}\}.}
    \label{fig:freqvar}
\end{figure}

Despite the curve shifts with temperature, if we consider the relative frequency variation, we can simplify the body bias model approximating the curve with a linear function having $5\%/\SI{100}{\milli \volt}$ (at 0.7V) as slope (\figurename{ \ref{fig:freqvar}}).
Table \ref{tab:perf_gain} shows the same analysis at different voltage operating points.

\begin{table}[tb]
    \centering
    \begin{tabular}{c|c|c}
    \toprule
        PULP \SI{0.5}{\volt} & \SI{0.7}{\volt} & \SI{0.9}{\volt}\\ \midrule
        $11\%/\SI{100}{\milli \volt}$ & $5\%/\SI{100}{\milli \volt}$& $3\%/\SI{100}{\milli \volt}$ \\
        \bottomrule
    \end{tabular}
    \\[5pt]
    \caption{Body bias induced performance gain}
    \label{tab:perf_gain}
\end{table}

This model allows to compute the additional body bias margin required to compensate the measurement errors described in the previous sections. Assuming to compensate the model uncertainty caused by temperature, which was the 3\% at \SI{0.7}{\volt}, we need to add a margin on $V_{BB}$ of \SI{75}{\milli \volt}.

It can be noted that body bias margins can be summed as independent contributions, to compensate different errors of the models. Assume to use a model which has several uncorrelated error sources: 3\% intrinsic error (PMB sensor), 1\% originated by temperature variations and 5.7\% generated by process variations. The resulting 9.7\% total error can be compensated with the sum of the body bias margins that would separately compensate each error\cite{Kumar2008a}.
As a concluding remark for this section, it can be noted that to minimize the error of the model, a calibration procedure capable to discover the chip process corner is required. This calibration procedure reduces the error by a factor of 2.4 and is described in section \ref{sec:controller}.


\subsection{Methodology Overhead}
\label{sec:overhead}


The compensation methodology we propose is based on the application of forward body bias to compensate process and temperature variations. As it is well known, FBB has a direct influence on the leakage current of a circuit, hence it is important to quantify the overhead associated with the additional body bias margins. Specifically, we defined the overhead as the ratio between the additional leakage caused by the body bias margin and the leakage current of the circuit when an ideal controller is regulating the body bias voltage (i.e. capable to apply the exact $V_{BB}$ to achieve a given target frequency). How the additional leakage current affects the global power consumption strongly depends on the dynamic power consumption (i.e operating frequency) of the circuit, hence the overhead will be computed with respect to the leakage current under the \textit{ideal} body bias regulation.


The overhead presented in this section refers to the three different models derived in the previous sections. More specifically, to the cases where the supply voltage is known and \textit{i)} the process corner and the temperature are unknown, we define this model as \textit{Process-Unaware Temperature-Unaware model} \textit{ii)} the process corner is known and the temperature is unknown, we refer to it as \textit{Process-Aware Temperature-Unaware model} \textit{iii)} both temperature and process corner are known, which is the \textit{Process-Aware Temperature-Aware model}.

When the \textit{Process-Unaware Temperature-Unaware} model is used, we experience the highest overhead. As expected, the higher is the uncertainty of the model, the wider is the body bias safety margin required to use the model, and also the higher is the associated overhead because of the higher leakage current. However, having the possibility to determine the chip process corner and calibrate the model on the specific device (i.e. passing to the \textit{Process-Aware Temperature-Unaware } model) the overhead can be significantly reduced. If we consider the leakage current associated with the body bias margins needed to correct the errors of the \textit{Process-Aware Temperature-Unaware} model, we can observe that the overhead is much lower. Finally, probing also the temperature, it would be possible to use multiple, better correlated, \textit{Process-Aware Temperature-Aware} PMB models at different temperatures. Table \ref{tab:temp-proc} summarizes the results in terms of overhead.

\begin{table}[hbt]
    \centering
    \begin{tabular}{l|c|c|c}
    \toprule
        $V_{dd}$ & \SI{0.9}{\volt} & \SI{0.7}{\volt} & \SI{0.5}{\volt}\\
        
    \midrule
        \multicolumn{4}{l}{Proc-unaware/Temp-unaware \hfill}\\
    \midrule
        $F_{err}$ & 6.6\% & 9.7\% & 25\% \\
        $V_{BB}$ margin [mV] & 150 &  150& 200\\
        Leakage overhead & 33\% & 37\% & 66\%\\
    \midrule
        \multicolumn{4}{l}{Proc-aware/Temp-unaware}\\
    \midrule
        $F_{err}$ & 3\% & 4\% & 7\% \\
        $V_{BB}$ margin [mV]& 100& 100& 100\\
        Leakage overhead & 13\% & 14\% & 15\%\\
    \midrule
        \multicolumn{4}{l}{Proc-aware/Temp-aware}\\
    \midrule
        $F_{err}$ & 2\% & 3\% & 6\% \\
        $V_{BB} $ margin [mV] & 50& 50& 50\\
        Leakage overhead & 9\% & 10\% & 12\%\\
   \bottomrule
    \end{tabular}
    \\[5pt]
    \caption{Frequency error of the three presented models and power consumption overheads with respect to the ideal compensation where no body bias margins are used.}
    \label{tab:temp-proc}
\end{table}

\section{MODEL UTILIZATION}
\label{sec:controller}
%
%

\subsection{Calibration Procedure}

As we have demonstrated \fix{}{it} in section \ref{sec:model}, the model derived from the PMB frequency estimations of chips belonging to different process corners is affected by a significant error. However, this error can be reduced by adopting a calibration procedure. More specifically, the operation performed during the characterization phase on the testing equipment for a single temperature operating point can be replicated on the embedded board. The constraint to ensure proper calibration is to execute the procedure at a \textcolor{black}{constant temperature. This goal can be achieved by exploiting an on-chip temperature sensor. For a correct calibration, the absolute value of the temperature at which the calibration is executed is not relevant, however the temperature must remain constant to not influence the characterization of the PMB sensor. Then, starting from a controller calibrated on a constant temperature value, two scenarios are possible:
\textit{i)} the temperature does not change with respect to the calibration point, and the controller has to compensate with VBB margins the 3\% of model uncertainty at 0.7V (Fig. 5). \textit{ii)} the temperature changes with respect to the calibration point, and the controller has to compensate a slightly higher model uncertainty, which takes into account the effects of temperature change, which is approximately 4\% at 0.7V (Fig. 6)}. 

Here we describe the calibration procedure. At the boot, the benchmark application mentioned in section \ref{sec:model} is loaded in the device memory by an external microcontroller connected through an SPI interface. Then the application is executed in a loop which increases the operating frequency at every iteration by 1MHz. The frequency is increased until the chip starts to fail, either returning wrong arithmetic results of \fix{}{completely} failing. This test is executed on multiple points covering the entire body bias range to correlate the maximum frequency of the chip with the PMB frequency estimation. 

\begin{figure}[t]
    \centering
    \includegraphics[width = \columnwidth]{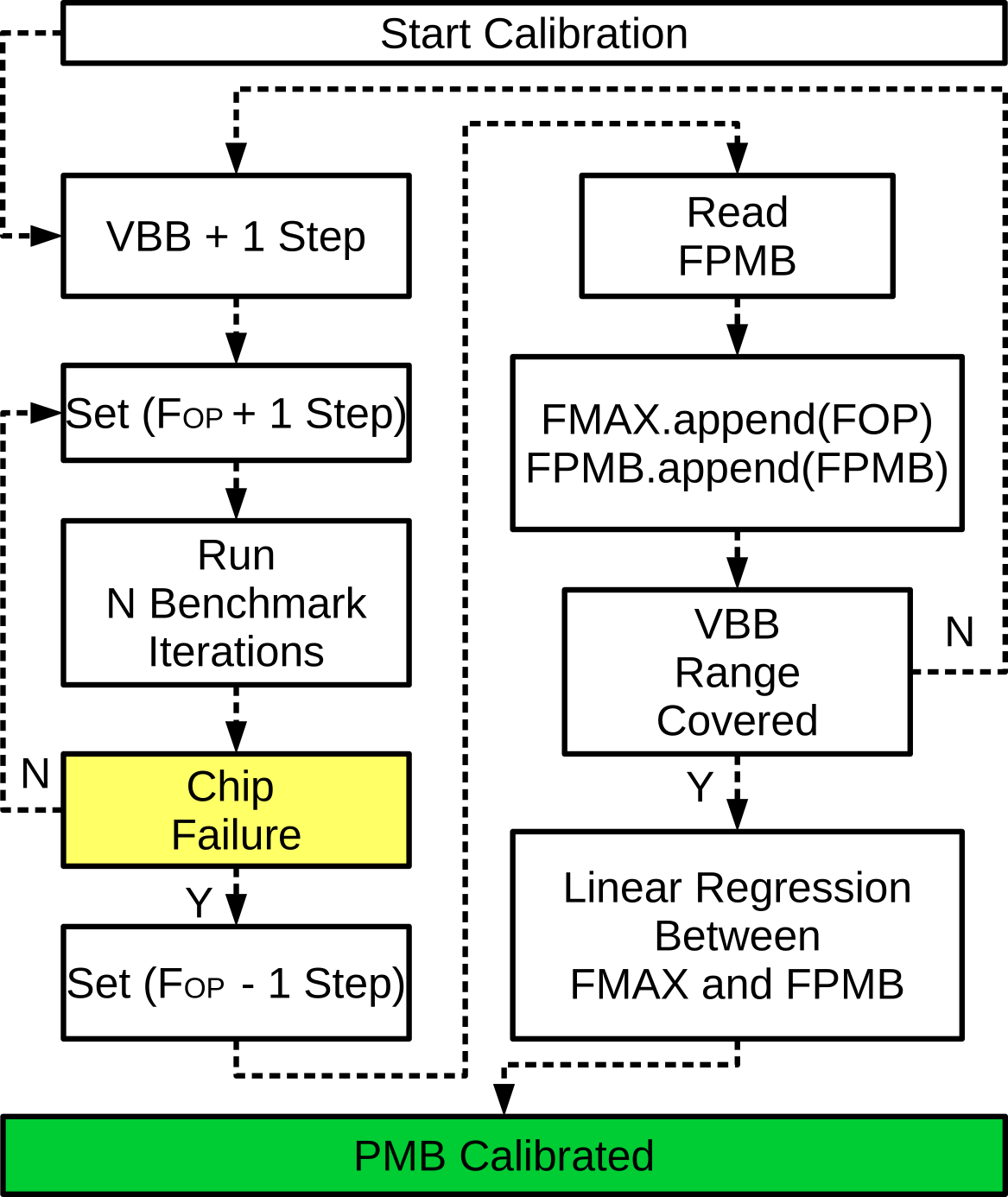}
    \caption{Calibration procedure executed on the embedded board to determine the process corner and tune the PMB model.}
    \label{fig:calibration}
\end{figure}

Using a model calibrated on the specific process corner, when $V_{DD}$ = \SI{0.7}{\volt}, it is possible to pass from the \textit{Process-unaware Temperature-unaware} to the \textit{Process-aware Temperature-unaware} (Table \ref{tab:temp-proc}), reducing the error by 2.4X. \textcolor{black}{The duration of a complete calibration at a single VDD operating point depends on the minimum step of the sweep performed on VBB and the maximum frequency. Additionally, the maximum frequency search method may increase the duration of the calibration procedure (i.e. linear sweep vs binary search). Finally the duration of the benchmark executed by the processor also changes the calibration time. In our example, we used a \SI{50}{\milli \volt} body bias voltage step to span the \SI{1.5}{\volt} range, resulting in 30 VBB points, and a linear sweep with a 1MHz step for the maximum frequency search. At every operating condition we executed 10000 benchmark iterations. The overall duration, starting from a \SI{100}{\mega \hertz} initial frequency for the maximum frequency search, lasted approximately 6 seconds.}
The procedure to perform this operation is illustrated as block diagram representation in Figure \ref{fig:calibration}.

\subsection{Body-Bias Controller}
In this section we will show how the models derived so far and selected with the procedure described in the previous section can be used in practical applications. More specifically, an accurate on-chip performance feedback, as well as the modeled behavior of the body-bias generator enable the building of software control systems; the structure we propose is a mixed Hardware-Software solution which properly set the cores cluster Body-Bias voltage depending on the clock frequency set-point provided as input.

\begin{figure}[tb]
    \centering
    \includegraphics[width = \columnwidth]{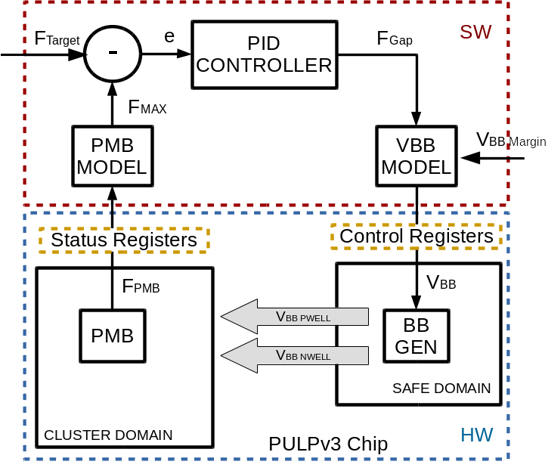}
    \caption{Block diagram representation of the body-bias controller.}
    \label{fig:control}
\end{figure}

\subsection{Controller Modules}
The proposed body-bias controller architecture is composed of: \textit{i)} a feedback, \textit{ii)} a subtractor which computes the mismatch between feedback and set-point, \textit{iii)} a Proportional-Integrative-Derivative (PID) controller, \textit{iv)} an actuator fed with the PID controller output (VBB MODEL + BB GEN). 
\figurename{ \ref{fig:control}} shows a block diagram of the body bias control system.

The control system \textit{feedback} module is divided in two main building blocks: \textit{i)} a hardware component, the previously mentioned PMB, which returns a raw maximum frequency estimation; \textit{ii)} the software model derived for the PMB, which allows to properly convert the raw output of the sensor in a clock frequency value. 

The \textit{actuator}, as in the case of feedback module, is composed of two building blocks: the on-chip Body-Bias generator, paired with a software model which allows to determine the right amount of Body-Bias needed to fill the frequency gap between feedback and set-point.

The \textit{PID controller} is a standard well-known control mechanism used in industrial control systems \cite{Wei1999}, the proposed body bias control system  is entirely implemented in software and it is fed with the mismatch between feedback and set-point frequency.
The PID controller has to be tuned accordingly to the feedback and the actuator, this process can be done empirically. In the context of this control task we decided to adopt a parameter set which minimizes the settling time while keeping under control undershoots. \textcolor{black}{Note that, for this kind of system, undershoots are more critical than overshoots. During an overshoot, the chip is biased with more than necessary FBB. Therefore, For a short amount of time, the maximum frequency of the chip is faster than than expected . On the contrary, during undershoots, the controller applies less FBB than required (or even RBB). Therefore, the actual maximum frequency of the chip can be lower than the requested frequency. This condition is critical and very likely determines a chip failure}



As shown in \figurename{ \ref{fig:control}}, the entry point of the control system is the frequency set-point $F_{Target}$. The first operation performed by the controller is to measure the maximum frequency of the system. Once the output of the PMB module is ready, it is converted in a frequency value, and it is compared with the input set-point obtaining the frequency mismatch. Then, the PID module is fed with the frequency mismatch and its output is sent to the body-bias software model. Finally, once the \fix{}{controller} computes the body-bias voltage to apply, it is used to set the body-bias generator to the new $V_{BB}$ value.

\subsection{Controller operation}

\subsubsection{Frequency Tracking}
\begin{figure}[tb]
    \centering
    \includegraphics[width = \columnwidth]{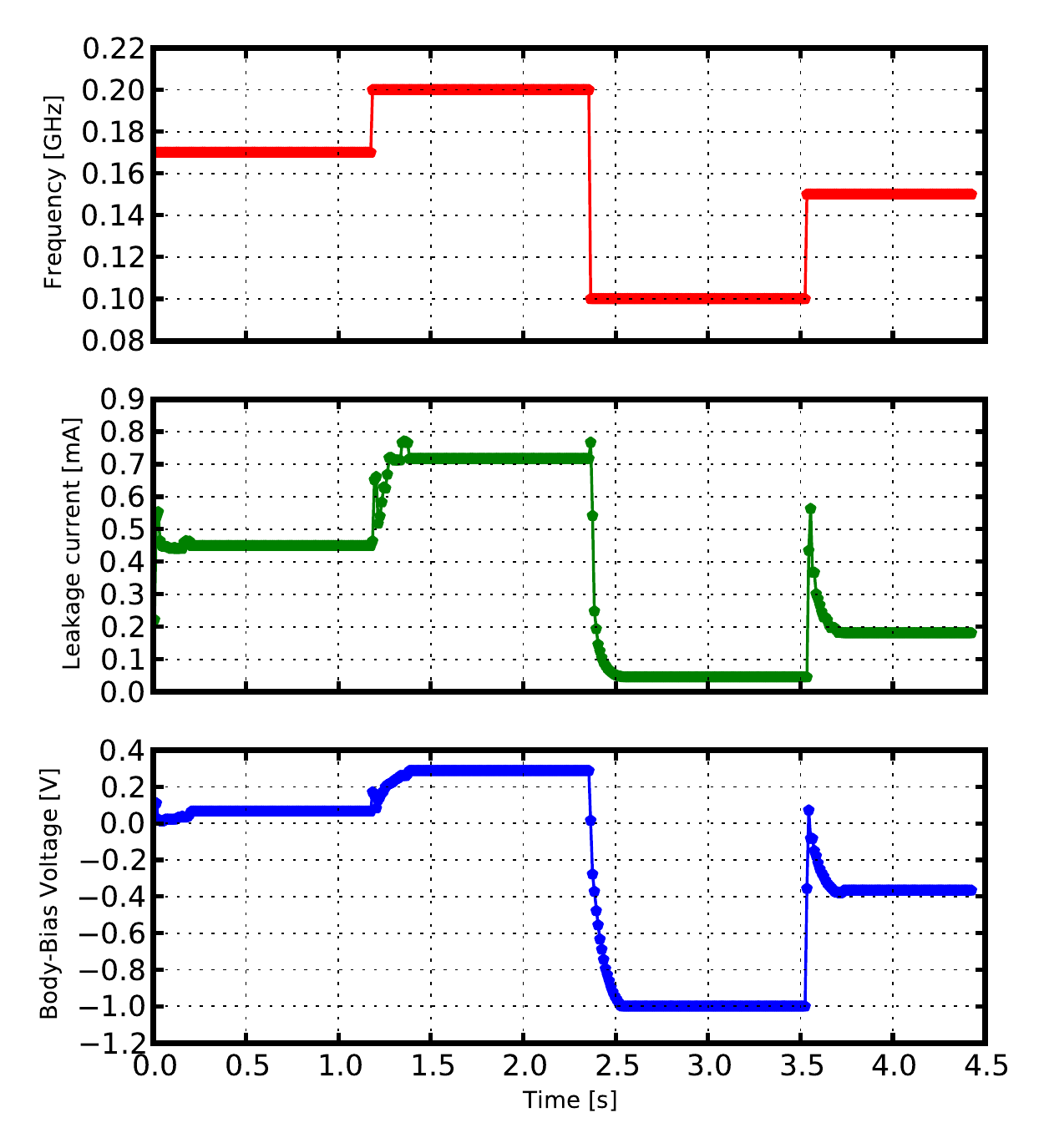}
    \caption{Body bias voltage regulation operated by the controller during boosting and leakage reduction phase, respectively. Supply voltage is 0.7V}
    \label{fig:control_op}
\end{figure}

In the following we present a controller working example. \figurename{ \ref{fig:control_op}} shows the $V_{BB}$ regulation operated by the controller when the frequency set-point is changed to different values \footnote{\textcolor{black}{Every time a new frequency is requested, the controller reset the body bias to VBB = \SI{0}{\volt} in a single step, and uses this voltage as a starting point. This choice has been taken to easy the implementation and reduce on the average the regulation time when changing from one frequency set-point to the next}}.

The first set-point, 175MHz, is very close \fix{}{to} the maximum frequency of the device. In this condition the controller \fix{}{applies} a very small amount of body-bias, since the system is already capable to run at the requested frequency, without additional forward body-bias. The only body bias voltage applied ($V_{BB} \neq 0$) is the body bias margin. When 200MHz are requested as set-point, the controller applies a forward body-bias voltage of approximately 300mV; applying a strong forward body bias, the system is now capable to run at 200MHz. On the contrary, when a slow frequency is set as set-point, 100MHz, the controller applies an aggressive reverse body-bias to reduce leakage power as much as possible. Finally, when the set-point is set to 150MHz, the system reduce the reverse body bias, applying to -400mV.

\subsubsection{Temperature tracking}

\begin{figure}[tb]
    \centering
    \includegraphics[width = \columnwidth]{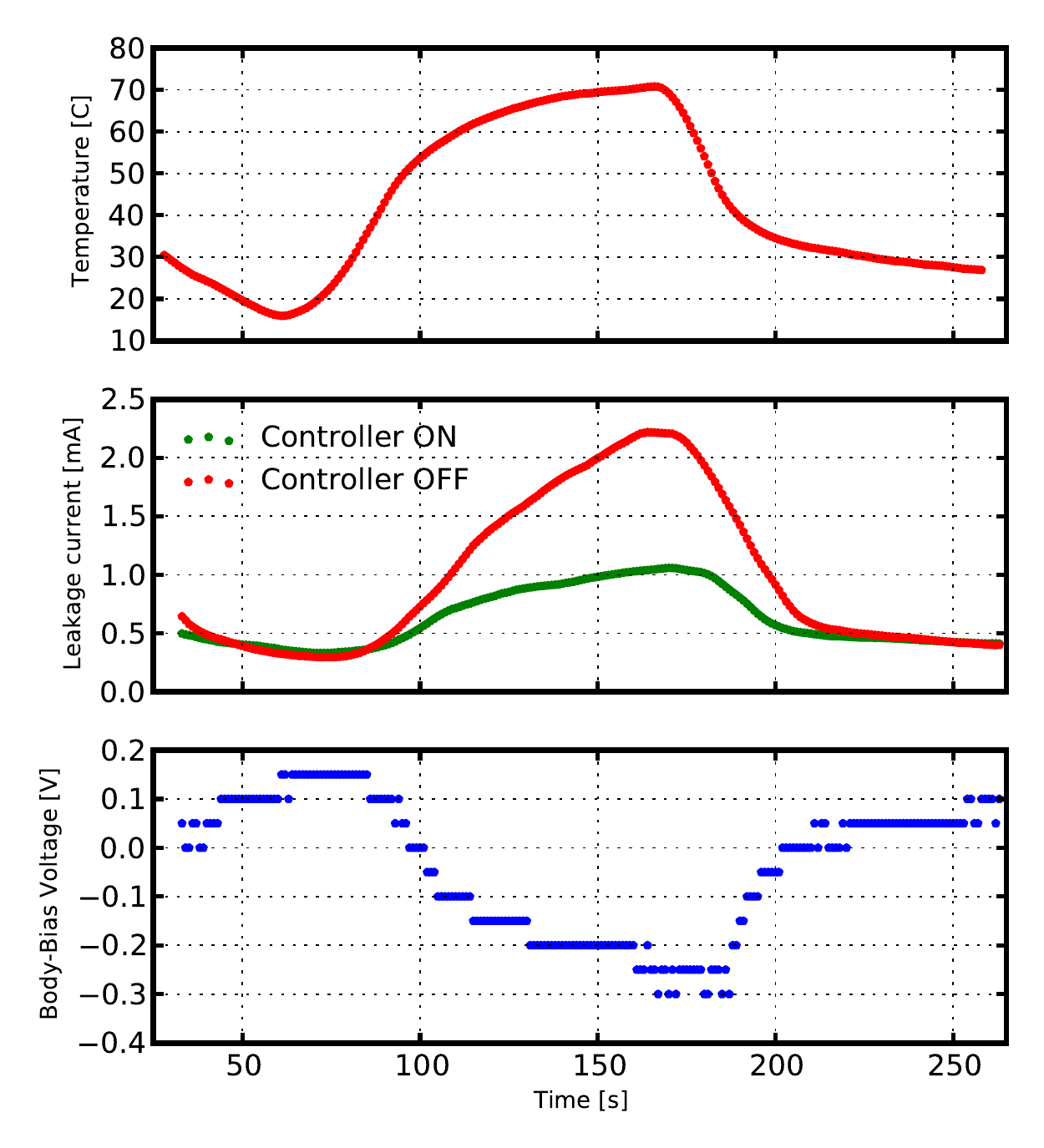}
    \caption{This plot shows in parallel the body-bias voltage modulation, the environmental temperature variation and the leakage current when the body bias controller is active.}
    \label{fig:controller_time}
\end{figure}

Here we show another working example. In this case the controller is used to guarantee a 170MHz frequency target against environmental temperature changes. \figurename{ \ref{fig:controller_time}} shows the behaviour of temperature, leakage and body-bias voltage when the controller is active.

From the leakage current plot it is clear that the controller is compensating the environmental temperature increase applying a reverse body-bias voltage. On the contrary, when the temperature decreases, to guarantee the requested target frequency the controller applies a forward body-bias voltage to boost the maximum performance of the chip.

\section{Results}
\label{sec:results}

In this section we present results in terms of energy efficiency gain and leakage reduction when the body-bias controller is turned on. 

\begin{figure}[h]
    \centering
    \includegraphics[width = \columnwidth]{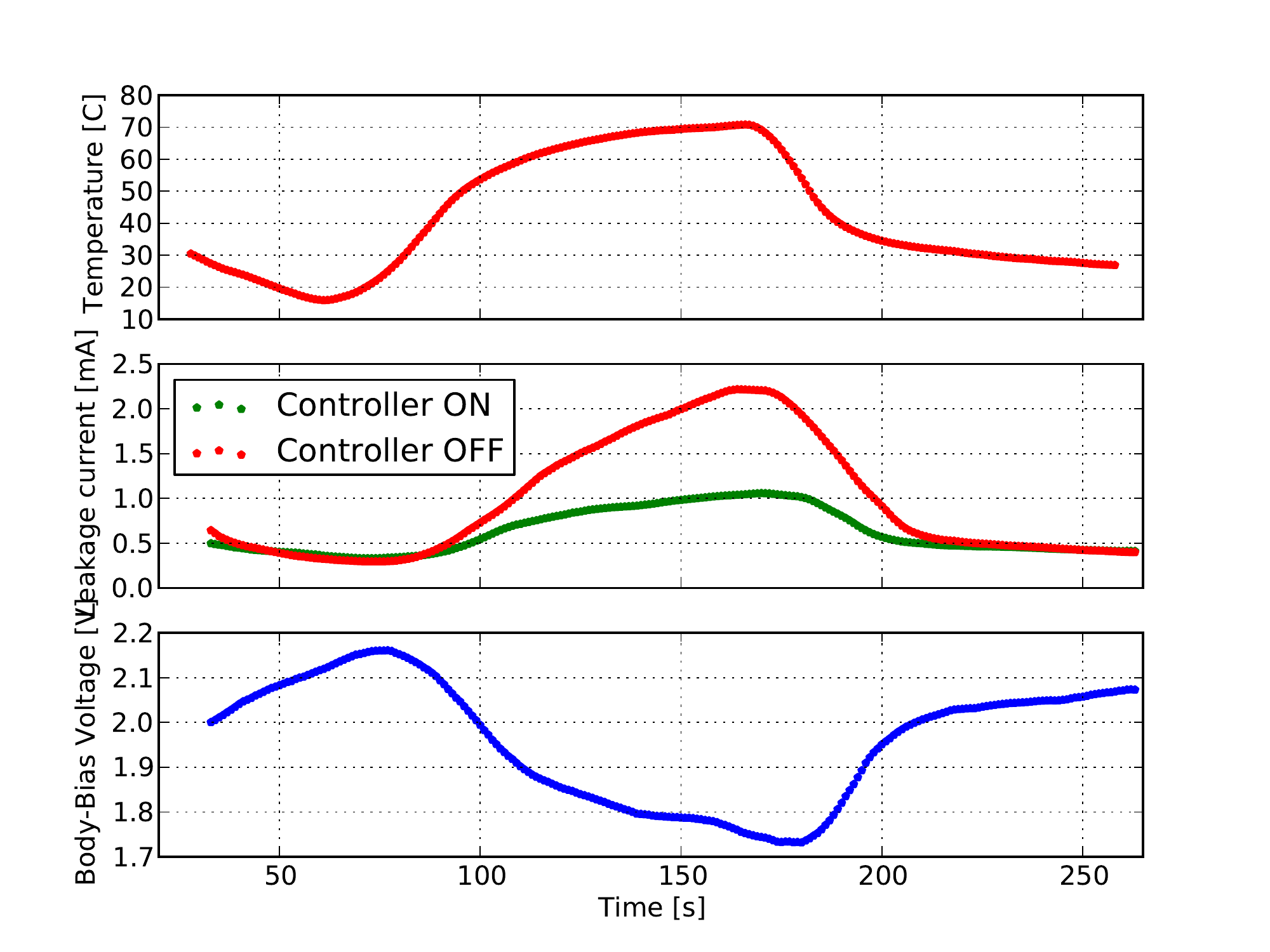}
    \caption{\fix{}{This plot shows the leakage current versus temperature, the operating frequency is 170MHz and the supply voltage is 0.7V. The red curve reports the leakage current when the controller is off. The green curve shows the Leakage current when the controller is on; the process corner is known. The yellow curve refers to the case where the controller is on but the process corner is not known. To guarantee the operating frequency, additional FBB margins are used, resulting in a leakage current increase. Blue curve shows an ideally compensated leakage current, i.e. when no FBB margins are used by the controller.} Note that below 17 \textdegree C \fix{}{it} is not possible to sustain the 170MHz target frequency without applying forward body biasing, hence the circuit stops to work.}
    \label{fig:leakage}
\end{figure}

\figurename{ \ref{fig:leakage}} shows a comparison between the leakage current when \textit{i)} the controller is off and the leakage is not compensated against temperature \textit{ii)} the controller is on and it is regulating the $V_{BB}$ using the \textit{Process-aware Temperature-unaware} model \textit{iii)} the controller is on and it is regulating the $V_{BB}$ using the \textit{Process-unaware Temperature-unaware} model
\textit{iv)} the ideal case where the leakage is compensated without additional body bias margins, hence assuming no error in the model converting the frequency estimation of the PMB. Note that the influence of the leakage power on the global power consumption depends on the operating frequency of the device, in our measurements the controller is regulating the body bias to achieve a target frequency of 170MHz, which represents an optimal operating point, since the working frequency is close to the maximum one.

\figurename{ \ref{fig:leakage}} also shows that the margin required to compensate process variations causes a big increase in the leakage current. In this context the effect on the global power consumption is not negligible. However this problem can be solved by running a simple calibration procedure which determines if the chip is a \textit{Slow}, \textit{Typical} or a \textit{Fast} one. 

The benefit of the calibration procedure is to reduce the error of the model, allowing to use the \textit{Process-aware Temperature-unaware} model instead of the \textit{Process-unaware Temperature-unaware}. As a consequence, the body bias margin to apply is reduced from 150mV to 100mV.

\begin{figure}[h]
    \centering
    \includegraphics[width = \columnwidth]{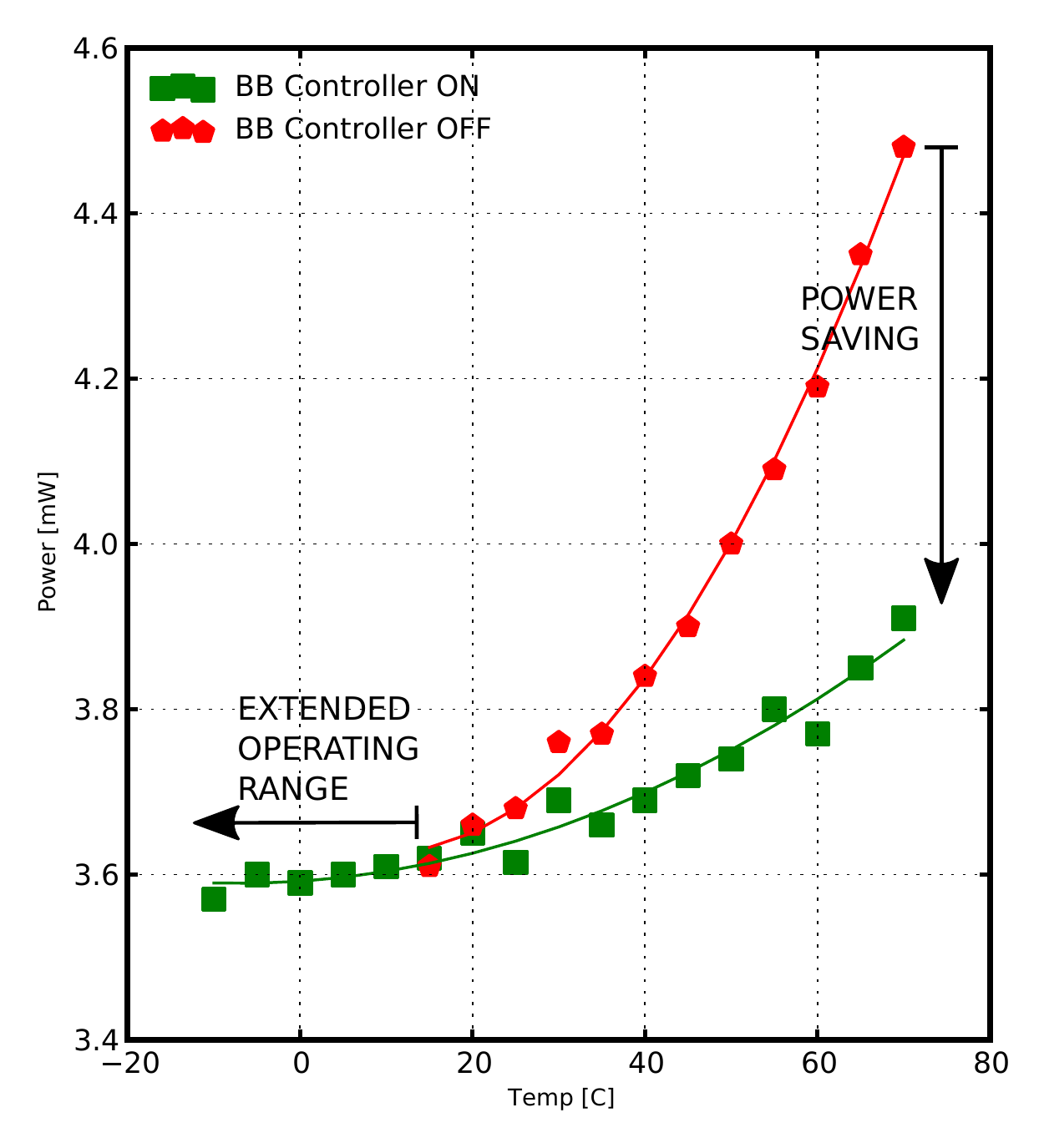}
    \caption{Comparison between power consumption with and without the performance aware body bias controller performed at 170MHz, $V_{DD}$ = \SI{0.7}{\volt}. Note that below 17 \textdegree C is not possible to run at 170MHz without compensating the performance degradation with additional body biasing (i.e. with the controller on).}
    \label{fig:power}
\end{figure}

\figurename{ \ref{fig:power}} compares the power consumption of the chip, at different temperatures, both in the case when the body bias controller is turned on and off. As it is evident from the plot, the use of the body bias controller allows to improve the energy efficiency by 15\% (at high temperatures), and extend the operating range to very low temperatures necessary to deal with the wide range of environmental conditions of IoT devices, leveraging a low-margins design methodology for energy efficient implementation.

\section{Conclusion}
\label{sec:conclusion}

In this work we demonstrated the body-biasing capabilities of the UTBB FD-SOI technology in compensating temperature and process variations.

We performed an analysis on Process Monitoring Blocks to correlate the on-chip estimated maximum frequency to the actual performance of the device, hence we derived a calibration model for the PMB sensors. We extended the model obtained at a single temperature to multiple temperatures, to compensate the effect that environmental temperature changes have on the performance of the device. Then, we further generalized the model to the situation where the process corner is unknown, considering it as an uncertainty of the model. We found that at \SI{0.7}{\volt} the frequency estimation error is in the order of 3\% when both process and temperature are determined, 4\% if the temperature is unknown and 9.7\% when also the process corner is unknown. We developed a strategy to eliminate such frequency estimation error by over-compensating with selective body bias margins. We also derived an on-board calibration procedure capable to determine the process corner of the chip and to select the proper \textit{Process-aware Temperature-unaware} model.
Once the models have been obtained, we developed a closed loop control strategy for the body bias voltage based on a simple PID controller. We tested the controller on a real embedded platform, demonstrating that with a minimum overhead, an automatic body bias regulation can reduce by a factor of 2 the leakage power consumption caused by temperature changes. Finally we evaluated the effect of the leakage reduction in a common operating context, demonstrating that our approach can introduce a \fix{}{15\%} energy efficiency improvement on the global power consumption.

\section*{Acknowledgment}
We thank STMicroelectronics for chip fabrication. This work was supported by European project ExaNoDe (H2020-671578) and European project OPRECOMP (732631).



\bibliographystyle{IEEEtran}
\bibliography{bibliography.bib}

\begin{thebibliography}{10}
\providecommand{\url}[1]{#1}
\csname url@samestyle\endcsname
\providecommand{\newblock}{\relax}
\providecommand{\bibinfo}[2]{#2}
\providecommand{\BIBentrySTDinterwordspacing}{\spaceskip=0pt\relax}
\providecommand{\BIBentryALTinterwordstretchfactor}{4}
\providecommand{\BIBentryALTinterwordspacing}{\spaceskip=\fontdimen2\font plus
\BIBentryALTinterwordstretchfactor\fontdimen3\font minus
  \fontdimen4\font\relax}
\providecommand{\BIBforeignlanguage}[2]{{%
\expandafter\ifx\csname l@#1\endcsname\relax
\typeout{** WARNING: IEEEtran.bst: No hyphenation pattern has been}%
\typeout{** loaded for the language `#1'. Using the pattern for}%
\typeout{** the default language instead.}%
\else
\language=\csname l@#1\endcsname
\fi
#2}}
\providecommand{\BIBdecl}{\relax}
\BIBdecl

\bibitem{Dreslinski2010a}
R.~G. Dreslinski, M.~Wieckowski, D.~Blaauw, D.~Sylvester, and T.~N. Mudge,
  ``Near-threshold computing: Reclaiming moore's law through energy efficient
  integrated circuits,'' \emph{Proceedings of the {IEEE}}, vol.~98, no.~2, pp.
  253--266, 2010.

\bibitem{Markovic2010}
D.~Markovic, C.~C. Wang, L.~P. Alarc{\'{o}}n, T.~Liu, and J.~M. Rabaey,
  ``Ultralow-power design in near-threshold region,'' \emph{Proceedings of the
  {IEEE}}, vol.~98, no.~2, pp. 237--252, 2010.

\bibitem{Rossi2017a}
D.~Rossi, A.~Pullini, I.~Loi, M.~Gautschi, F.~K. G{\"{u}}rkaynak, A.~Teman,
  J.~Constantin, A.~Burg, I.~Miro-panades, E.~Beign{\`{e}}, P.~Flatresse, and
  L.~Benini, ``{Near-Threshold Parallel Computing : The PULPv2 Cluster},''
  2017.

\bibitem{Pahlevan2016}
A.~Pahlevan, J.~Picorel, A.~P. Zarandi, D.~Rossi, M.~Zapater, A.~Bartolini,
  P.~G.~D. Valle, D.~Atienza, L.~Benini, and B.~Falsafi, ``Towards
  near-threshold server processors,'' in \emph{2016 Design, Automation Test in
  Europe Conference Exhibition (DATE)}, March 2016, pp. 7--12.

\bibitem{Alioto2017}
M.~Alioto, Ed., \emph{Enabling the Internet of Things}.\hskip 1em plus 0.5em
  minus 0.4em\relax Springer International Publishing, 2017.

\bibitem{Lee2014}
\BIBentryALTinterwordspacing
W.~Lee, Y.~Wang, T.~Cui, S.~Nazarian, and M.~Pedram, ``Dynamic thermal
  management for {FinFET}-based circuits exploiting the temperature effect
  inversion phenomenon,'' in \emph{Proceedings of the 2014 International
  Symposium on Low Power Electronics and Design}, ser. ISLPED '14.\hskip 1em
  plus 0.5em minus 0.4em\relax New York, NY, USA: ACM, 2014, pp. 105--110.
  [Online]. Available: \url{http://doi.acm.org/10.1145/2627369.2627608}
\BIBentrySTDinterwordspacing

\bibitem{Pu2010}
Y.~Pu, X.~Zhang, J.~Huang, A.~Muramatsu, M.~Nomura, and K.~Hirairi,
  ``{Misleading Energy and Performance Claims in Sub / Near Threshold Digital
  Systems},'' \emph{Technology}, pp. 4--6, 2010.

\bibitem{Han2017}
\BIBentryALTinterwordspacing
K.~Han, J.-J. Lee, J.~Lee, W.~Lee, and M.~Pedram, ``{TEI-NoC: Optimizing
  Ultra-Low Power NoCs Exploiting the Temperature Effect Inversion},''
  \emph{IEEE Transactions on Computer-Aided Design of Integrated Circuits and
  Systems}, vol. 0070, no.~c, pp. 1--1, 2017. [Online]. Available:
  \url{http://ieeexplore.ieee.org/document/7896585/}
\BIBentrySTDinterwordspacing

\bibitem{Alioto2012a}
M.~Alioto, ``{Ultra-Low Power {\{}VLSI{\}} Circuit Design Demystified and
  Explained: A Tutorial},'' \emph{{\{}IEEE{\}} Transactions on Circuits and
  Systems I: Regular Papers}, vol.~59, no.~1, pp. 3--29, jan 2012.

\bibitem{Sundaresan}
K.~Sundaresan, K.~Brouse, K.~U-Yen, F.~Ayazi, and P.~Allen, ``A 7-{MHz}
  process, temperature and supply compensated clock oscillator in 0.25 $\mu$m
  {CMOS},'' in \emph{Proceedings of the 2003 International Symposium on
  Circuits and Systems, 2003. {ISCAS} {\textquotesingle}03.}\hskip 1em plus
  0.5em minus 0.4em\relax {IEEE}, 2003.

\bibitem{Gammie2008}
G.~Gammie, A.~Wang, M.~Chau, S.~Gururajarao, R.~Pitts, F.~Jumel, S.~Engel,
  P.~Royannez, R.~Lagerquist, H.~Mair, J.~Vaccani, G.~Baldwin, K.~Heragu,
  R.~Mandal, M.~Clinton, D.~Arden, and U.~Ko, ``A 45nm 3.5g
  baseband-and-multimedia application processor using adaptive body-bias and
  ultra-low-power techniques,'' in \emph{2008 {IEEE} International Solid-State
  Circuits Conference - Digest of Technical Papers}.\hskip 1em plus 0.5em minus
  0.4em\relax {IEEE}, feb 2008.

\bibitem{Clerc2015}
S.~Clerc, M.~Saligane, F.~Abouzeid, M.~Cochet, J.-M. Daveau, C.~Bottoni,
  D.~Bol, J.~De-Vos, D.~Zamora, B.~Coeffic \emph{et~al.}, ``8.4 a 0.33 v/-40°
  c process/temperature closed-loop compensation soc embedding all-digital
  clock multiplier and dc-dc converter exploiting fdsoi 28nm back-gate
  biasing,'' in \emph{Solid-State Circuits Conference-(ISSCC), 2015 IEEE
  International}.\hskip 1em plus 0.5em minus 0.4em\relax IEEE, 2015, pp. 1--3.

\bibitem{Rossi2015a}
D.~Rossi, A.~Pullini, M.~Gautschi, I.~Loi, F.~K. Gurkaynak, P.~Flatresse, and
  L.~Benini, ``A 60 gops/w,-1.8 v to 0.9 v body bias ulp cluster in 28nm utbb
  fd-soi technology,'' in \emph{2015 {IEEE} {SOI}-3D-Subthreshold
  Microelectronics Technology Unified Conference (S3S)}.\hskip 1em plus 0.5em
  minus 0.4em\relax {IEEE}, oct 2015.

\bibitem{Rossi2017}
D.~Rossi, I.~Loi, F.~Conti, L.~Benini, C.~M{\"{u}}ller, and A.~Burg, ``{A
  Self-Aware Architecture for PVT Compensation and Power Nap in Near- Threshold
  Processors},'' pp. 46--53, 2017.

\bibitem{Seok2011}
M.~Seok, G.~K. Chen, S.~Hanson, M.~Wieckowski, D.~T. Blaauw, and D.~Sylvester,
  ``{{\{}CAS-FEST{\}} 2010: Mitigating Variability in Near-Threshold
  Computing},'' \emph{{\{}IEEE{\}} J. Emerg. Sel. Topics Circuits Syst.},
  vol.~1, no.~1, pp. 42--49, 2011.

\bibitem{Ernst2003}
D.~Ernst, N.~S. Kim, S.~Das, S.~Pant, R.~Rao, T.~Pham, C.~Ziesler, D.~Blaauw,
  T.~Austin, K.~Flautner, and T.~Mudge, ``{Razor: A low-power pipeline based on
  circuit-level timing speculation},'' \emph{Proceedings of the Annual
  International Symposium on Microarchitecture, MICRO}, vol. 2003-January, pp.
  7--18, 2003.

\bibitem{Das2005}
S.~Das, S.~Pant, D.~Roberts, S.~Lee, D.~Blaauw, T.~Austin, T.~Mudge, and
  K.~Flautner, ``{A self-tuning DVS processor using delay-error detection and
  correction},'' \emph{IEEE Symposium on VLSI Circuits, Digest of Technical
  Papers}, vol. 2005, no.~4, pp. 258--261, 2005.

\bibitem{Blaauw_2008}
D.~Blaauw, S.~Kalaiselvan, K.~Lai, W.-H. Ma, S.~Pant, C.~Tokunaga, S.~Das, and
  D.~Bull, ``Razor {II}: In situ error detection and correction for {PVT} and
  {SER} tolerance,'' in \emph{2008 {IEEE} International Solid-State Circuits
  Conference - Digest of Technical Papers}.\hskip 1em plus 0.5em minus
  0.4em\relax {IEEE}, feb 2008.

\bibitem{Bull2011}
D.~Bull, S.~Das, K.~Shivashankar, G.~S. Dasika, K.~Flautner, and D.~Blaauw,
  ``{A Power-Efficient 32 bit {\{}ARM{\}} Processor Using Timing-Error
  Detection and Correction for Transient-Error Tolerance and Adaptation to
  {\{}PVT{\}} Variation},'' \emph{{\{}IEEE{\}} Journal of Solid-State
  Circuits}, vol.~46, no.~1, pp. 18--31, jan 2011.

\bibitem{Fojtik2013}
M.~Fojtik, D.~Fick, Y.~Kim, N.~Pinckney, D.~M. Harris, D.~Blaauw, and
  D.~Sylvester, ``{Bubble razor: Eliminating timing margins in an ARM cortex-M3
  Processor in 45 nm CMOS using architecturally independent error detection and
  correction},'' \emph{IEEE Journal of Solid-State Circuits}, vol.~48, no.~1,
  pp. 66--81, 2013.

\bibitem{Drake2007}
A.~Drake, R.~Senger, H.~Deogun, G.~Carpenter, S.~Ghiasi, T.~Nguyen, N.~James,
  M.~Floyd, and V.~Pokala, ``{A distributed critical-path timing monitor for a
  65nm high-performance microprocessor},'' \emph{Digest of Technical Papers -
  IEEE International Solid-State Circuits Conference}, 2007.

\bibitem{Tschanz2009}
J.~Tschanz, K.~Bowman, S.~Walstra, M.~Agostinelli, T.~Karnik, and V.~De,
  ``{Tunable replica circuits and adaptive voltage-frequency techniques for
  dynamic voltage, temperature, and aging variation tolerance},'' \emph{2009
  Symposium on VLSI Circuits}, pp. 112--113, 2009.

\bibitem{Mhz2015}
A.~Mhz, V.~Ghz, E.~Beign{\'{e}}, A.~Valentian, I.~Miro-panades, R.~Wilson,
  P.~Flatresse, F.~Abouzeid, T.~Benoist, C.~Bernard, S.~Bernard, O.~Billoint,
  S.~Clerc, B.~Giraud, A.~Grover, J.~L. Coz, J.-p. Noel, O.~Thomas, and
  Y.~Thonnart, ``A 460 mhz at 397 mv, 2.6 ghz at 1.3 v, 32 bits vliw dsp
  embedding fmax tracking,'' vol.~50, no.~1, pp. 125--136, 2015.

\bibitem{Zandrahimi2016}
M.~Zandrahimi, Z.~Al-Ars, P.~Debaudand, and A.~Castillejo, ``Challenges of
  using on-chip performance monitors for process and environmental variation
  compensation,'' in \emph{Proceedings of the 2016 Design, Automation {\&} Test
  in Europe Conference {\&} Exhibition ({DATE})}.\hskip 1em plus 0.5em minus
  0.4em\relax Research Publishing Services, 2016.

\bibitem{Constantin2016}
J.~Constantin, A.~Bonetti, A.~Teman, C.~M{\"{u}}ller, L.~Schmid, and A.~Burg,
  ``{DynOR: A 32-bit microprocessor in 28 nm FD-SOI with cycle-by-cycle dynamic
  clock adjustment},'' \emph{European Solid-State Circuits Conference}, vol.
  2016-October, pp. 261--264, 2016.

\bibitem{Bowman2011}
K.~A. Bowman, J.~W. Tschanz, S.~L.~L. Lu, P.~A. Aseron, M.~M. Khellah,
  A.~Raychowdhury, B.~M. Geuskens, C.~Tokunaga, C.~B. Wilkerson, T.~Karnik, and
  V.~K. De, ``{A 45 nm resilient microprocessor core for dynamic variation
  tolerance},'' \emph{IEEE Journal of Solid-State Circuits}, vol.~46, no.~1,
  pp. 194--208, 2011.

\bibitem{Miyazaki}
M.~Miyazaki, G.~Ono, T.~Hattori, K.~Shiozawa, K.~Uchiyama, and K.~Ishibashi,
  ``A 1000-{MIPS}/w microprocessor using speed adaptive threshold-voltage
  {CMOS} with forward bias,'' in \emph{2000 {IEEE} International Solid-State
  Circuits Conference. Digest of Technical Papers (Cat. No.00CH37056)}.\hskip
  1em plus 0.5em minus 0.4em\relax {IEEE}, 2000.

\bibitem{Tschanz2002}
J.~W. Tschanz, J.~T. Kao, S.~G. Narendra, R.~Nair, D.~A. Antoniadis, A.~P.
  Chandrakasan, and V.~De, ``{Adaptive body bias for reducing impacts of
  die-to-die and within-die parameter variations on microprocessor frequency
  and leakage},'' \emph{IEEE Journal of Solid-State Circuits}, vol.~37, no.~11,
  pp. 1396--1402, 2002.

\bibitem{Kumar2008a}
S.~Kumar, C.~Kim, and S.~Sapatnekar, ``Body bias voltage computations for
  process and temperature compensation,'' \emph{{IEEE} Transactions on Very
  Large Scale Integration ({VLSI}) Systems}, vol.~16, no.~3, pp. 249--262, mar
  2008.

\bibitem{Tschanz2007}
J.~Tschanz, N.~Kim, S.~Dighe, J.~Howard, G.~Ruhl, S.~R. Vangal, S.~Narendra,
  Y.~Hoskote, H.~Wilson, C.~Lam, M.~Shuman, C.~Tokunaga, D.~Somasekhar,
  S.~Tang, D.~Finan, T.~Karnik, N.~Borkar, N.~A. Kurd, and V.~De, ``Adaptive
  frequency and biasing techniques for tolerance to dynamic temperature-voltage
  variations and aging,'' in \emph{2007 {IEEE} International Solid-State
  Circuits Conference, {ISSCC} 2007, Digest of Technical Papers, San Francisco,
  CA, USA, February 11-15, 2007}.\hskip 1em plus 0.5em minus 0.4em\relax
  {IEEE}, 2007, pp. 292--604.

\bibitem{Kang2010}
K.~Kang, S.~P. Park, K.~Kim, and K.~Roy, ``{On-chip variability sensor using
  phase-locked loop for detecting and correcting parametric timing failures},''
  \emph{IEEE Transactions on Very Large Scale Integration (VLSI) Systems},
  vol.~18, no.~2, pp. 270--280, 2010.

\bibitem{Kumar2011}
S.~V. Kumar, C.~H. Kim, and S.~S. Sapatnekar, ``{Adaptive techniques for
  overcoming performance degradation due to aging in CMOS circuits},''
  \emph{IEEE Transactions on Very Large Scale Integration (VLSI) Systems},
  vol.~19, no.~4, pp. 603--614, 2011.

\bibitem{Ono}
G.~Ono, M.~Miyazaki, H.~Tanaka, N.~Ohkubo, and T.~Kawahara, ``Temperature
  referenced supply voltage and forward-body-bias control ({TSFC}) architecture
  for minimum power consumption [ubiquitous computing processors],'' in
  \emph{Proceedings of the 30th European Solid-State Circuits
  Conference}.\hskip 1em plus 0.5em minus 0.4em\relax {IEEE}, 2004.

\bibitem{Flatresse2013}
P.~Flatresse, B.~Giraud, J.~P. Noel, B.~Pelloux-Prayer, F.~Giner, D.~K. Arora,
  F.~Arnaud, N.~Planes, J.~L. Coz, O.~Thomas, S.~Engels, G.~Cesana, R.~Wilson,
  and P.~Urard, ``{Ultra-wide body-bias range LDPC decoder in 28nm UTBB FDSOI
  technology},'' \emph{Digest of Technical Papers - IEEE International
  Solid-State Circuits Conference}, vol.~56, no. 424, pp. 424--425, 2013.

\bibitem{Conti2015}
F.~Conti, D.~Rossi, A.~Pullini, I.~Loi, and L.~Benini, ``{PULP}: A ultra-low
  power parallel accelerator for energy-efficient and flexible embedded
  vision,'' \emph{Journal of Signal Processing Systems}, vol.~84, no.~3, pp.
  339--354, nov 2015.

\bibitem{Blagojevic2016}
M.~Blagojevic, M.~Cochet, B.~Keller, P.~Flatresse, A.~Vladimirescu, and
  B.~Nikolic, ``{A fast, flexible, positive and negative adaptive body-bias
  generator in 28nm FDSOI},'' \emph{IEEE Symposium on VLSI Circuits, Digest of
  Technical Papers}, vol. 2016-Septe, pp. 9--10, 2016.

\bibitem{Beigne2015}
E.~Beigne, A.~Valentian, I.~Miro-Panades, R.~Wilson, P.~Flatresse, F.~Abouzeid,
  T.~Benoist, C.~Bernard, S.~Bernard, O.~Billoint, S.~Clerc, B.~Giraud,
  A.~Grover, J.~L. Coz, J.-P. Noel, O.~Thomas, and Y.~Thonnart, ``A 460 {MHz}
  at 397 {mV}, 2.6 {GHz} at 1.3 v, 32 bits {VLIW} {DSP} embedding f {MAX}
  tracking,'' \emph{{IEEE} Journal of Solid-State Circuits}, vol.~50, no.~1,
  pp. 125--136, jan 2015.

\bibitem{Mackay2006}
\BIBentryALTinterwordspacing
R.~S. Mackay, H.~Kamberian, and Y.~Zhang, ``{Methods to reduce lithography
  costs with reticle engineering},'' \emph{Microelectronic Engineering},
  vol.~83, no. 4-9, pp. 914--918, 2006. [Online]. Available:
  \url{http://linkinghub.elsevier.com/retrieve/pii/S0167931706001742}
\BIBentrySTDinterwordspacing

\bibitem{Wei1999}
G.~Y. Wei and M.~Horowitz, ``{Fully digital, energy-efficient, adaptive
  power-supply regulator},'' \emph{IEEE Journal of Solid-State Circuits},
  vol.~34, no.~4, pp. 520--528, 1999.

\end{thebibliography}

\end{document}